\documentclass[12pt,preprint]{aastex}
\usepackage{natbib,graphics,graphicx}

\slugcomment{\today}

\shorttitle{Quasars Unification through CIV Emission}
\shortauthors{Richards et al.}

\begin{document}

\newcommand{\civ}{\ion{C}{4}}
\newcommand{\mgii}{\ion{Mg}{2}}

\title{Unification of Luminous Type 1 Quasars through CIV Emission}

\author{
Gordon T. Richards,\altaffilmark{1,2}
Nicholas E. Kruczek,\altaffilmark{1}
S.~C. Gallagher,\altaffilmark{3}
Patrick B. Hall,\altaffilmark{4}
Paul C. Hewett,\altaffilmark{5}
Karen M. Leighly,\altaffilmark{6}
Rajesh P. Deo,\altaffilmark{1}
Rachael M. Kratzer,\altaffilmark{1}
and
Yue Shen\altaffilmark{7}
}

\altaffiltext{1}{Department of Physics, Drexel University, 3141 Chestnut Street, Philadelphia, PA 19104.}
\altaffiltext{2}{Alfred P. Sloan Research Fellow.}
\altaffiltext{3}{Department of Physics \& Astronomy, The University of Western Ontario, London, ON N6A 3K7, Canada.}
\altaffiltext{4}{Department of Physics and Astronomy, York University, Toronto, Ontario M3J 1P3, Canada.}
\altaffiltext{5}{Institute of Astronomy, Madingley Road, Cambridge CB3 0HA, UK.}
\altaffiltext{6}{Homer L. Dodge Department of Physics and Astronomy, The University of Oklahoma, 440 W. Brooks St., Norman, OK 73019.}
\altaffiltext{7}{Harvard-Smithsonian Center for Astrophysics, 60 Garden Street, MS-51, Cambridge, MA 02138.}

\begin{abstract}

  Using a sample of $\sim$30,000 quasars from the 7th Data Release of the Sloan Digital Sky Survey, we explore the range of properties exhibited by high-ionization, broad emission lines, such as \civ\ $\lambda1549$. Specifically we investigate the anti-correlation between continuum luminosity and emission line equivalent width (the Baldwin Effect) and the ``blueshifting'' of the high-ionization emission lines with respect to low-ionization emission lines.  Employing improved redshift determinations from Hewett \& Wild, the blueshift of the \civ\ emission line is found to be nearly ubiquitous, with a mean shift of $\sim810\,{\rm km\,s}^{-1}$ for radio-quiet quasars and $\sim360\,{\rm km\,s}^{-1}$ for radio-loud quasars.  The Baldwin Effect is present in both radio-quiet and radio-loud samples.  We consider these phenomena within the context of an accretion disk wind model that is modulated by the non-linear correlation between ultraviolet and X-ray continuum luminosity.  Composite spectra are constructed as a function of \civ\ emission line properties in attempt to reveal empirical relationships between different line species and the continuum.  Within a two-component disk+wind model of the broad emission line region (BELR), where the wind filters the continuum seen by the disk component, we find that radio-loud quasars are consistent with being dominated by the disk component, while broad absorption line quasars are consistent with being dominated by the wind component.  Some radio-quiet objects have emission line features similar to radio-loud quasars; they may simply have insufficient black hole spin to form radio jets.  Our results suggest that there could be significant systematic errors in the determination of $L_{\rm bol}$ and black hole mass that make it difficult to place these findings in a more physical context.  However, it is possible to classify quasars in a paradigm where the diversity of BELR parameters are due to differences in an accretion disk wind between quasars (and over time); these differences are underlain primarily by the spectral energy distribution, which ultimately must be tied to black hole mass and accretion rate.

\end{abstract}


\section{Introduction}

The unified model for active galactic nuclei (AGNs;
\citealt{ant93,up95}) has been sufficiently successful that it is
tempting to think of luminous quasars as being well described by a
static model in which the only free parameters are the presence of
radio jets or not and the angle of our line of sight to the central
engine.  Indeed, in the average sense, quasar spectra {\em are}
sufficiently similar that the use of composite spectra
\citep[e.g.,][]{fhf+91,vrb+01} can be very effective.  Nevertheless,
there is a large range of continuum, emission, and absorption
properties among quasars, which demands that quasars cannot be fully
described by a single, static picture.  Quasars are not {\em things}
so much as {\em processes} --- they are what happens when the central
black hole at the centers of massive galaxies are actively accreting
new material \citep[e.g.,][]{lb69}.  Moreover, it also seems that the
quasar stage is self-regulating, if not also of key importance to the
evolution of their host galaxy \citep[e.g.,][]{hhc+06}.

In addition to the rather obvious dichotomy between those objects
with strong radio jets and those without \citep[e.g.,][]{kss+89} and
the differences that result from different orientations of said jets,
there is a rich literature describing the more subtle differences
between AGNs.  We frequently carve out non-parametric classes such as
narrow line Seyfert 1s (NLS1s; \citealt{op85}) and broad absorption
line quasars (BALQSOs; \citealt{wmf+91}) to highlight extrema in
emission and absorption properties.  On a more subtle parametric
level, we can further quantify the uniqueness of each individual
quasar.  At low-redshift the parameters that best characterize the
diversity of broad emission line region (BELR) properties are the FWHM
of the H$\beta$ emission line and the strength of optical \ion{Fe}{2}
with respect to H$\beta$ (with [\ion{O}{3}] being anti-correlated with
\ion{Fe}{2}).  These parameters were identified as part of the
landmark study of \citet[hereafter BG92]{bg92} using a principal
component analysis (PCA) to identify those features that cause the
largest variance in the spectra of quasars.  The first two
eigenvectors of this PCA decomposition are generally referred to as
``eigenvector 1'' (EV1) and ``eigenvector 2'' (EV2).  By way of
comparison with Galactic black hole binaries \citep{pdo95,bbf96}, BG92
suggested that the Eddington ratio, $L_{\rm bol}/L_{\rm Edd}$, is the
primary driver of EV1.
Since that time \citet{wlb+99}, \citet{Laor2000a}, and \citet{szm+00},
among others, have added additional parameters (notably the X-ray
spectral index) to the eigenvector matrix and have extended the
analysis to higher redshifts.  In addition, {\em spectral} PCA
analyses \citep{fhf+92,Shang03,Yip04} have identified other
correlations between emission lines, broad absorption lines, and the
continuum.

At high-redshift, two well-known properties of the \civ\ emission line
are evident in quasar spectra (see Section~\ref{sec:c4prop}).  An
anti-correlation between luminosity and the equivalent width (EQW) of
\civ, generally known as the Baldwin Effect (BEff;
\citealt{Baldwin77}), was first identified in 20 (mostly radio-loud)
quasars and has been confirmed in much larger samples
\citep[e.g.,][]{wvb+09}.  The other effect is the blueshifting of the
\civ\ with respect to the systemic redshift \citep{gas82,Wilkes84}.
As with the BEff, these blueshifts have not only been confirmed with
larger samples, but appear to be nearly ubiquitous in the quasar
population \citep{szm+00,rvr+02}.

Complementing the range of emission line parameters in quasars is a
well-known aspect of quasar continua.  Specifically, there is a
non-linear scaling of $2\,{\rm keV}$ X-ray luminosity with the ultraviolet (UV)
luminosity at 2500\,\AA \citep[e.g.,][]{at82,gsa+95,ssb+06,jbs+07}.
This ``$L_{\rm UV}$--$\alpha_{\rm ox}$'' relationship is such that
quasars more luminous at UV wavelengths are (relatively) weaker at
X-ray (ionizing) wavelengths.

Considerable effort has been expended trying to explain/understand the
differences in emission-line and continuum properties of quasars.
With regard to the BEff, it is clear that the \civ\ emission line can
only be as strong as the number of ionizing photons available to
create triply-ionized carbon.  As such, the spectral energy
distribution (SED), in the form of the previously mentioned weakening
of the ionizing continuum with luminosity can be considered as the
``origin'' of the effect
\citep[e.g.,][]{ms82,nlg92,zm93,kbf98,Scott2004}.  Support for this
interpretation comes from evidence that the effect disappears when
referenced to the ionizing luminosity instead of the UV luminosity
\citep{Green96} and that there is an observed dependence on ionization
potential \citep[e.g.,][]{dhs+02}.  The ongoing debate regarding
whether the BEff is driven by the SED, black hole mass, accretion rate, or
Eddington ratio
\citep[e.g.,][]{nlg92,Wandel99,szm+00,bl04,bms+04,whd04,xby+08,wvb+09}
can really be thought of as a debate over the underlying cause of the
range of quasar SEDs as characterized by the ``$L_{\rm
  UV}$--$\alpha_{\rm ox}$'' relationship.

If the SED also influences the \civ\ blueshift, then it is important
to discuss the BEff and \civ\ blueshifts in the same context (and
together with the $L_{\rm UV}$--$\alpha_{\rm ox}$ relationship).
Ironically, the otherwise rather complete review of the BEff by
\citet{Shields07} fails to mention the line shifts at all, despite the
fact that large \civ\ blueshifts are known to be accompanied by weaker
lines \citep{cb96,rvr+02}.  Similarly, our analysis of the line shifts
in \citet{rvr+02} only considers the BEff as an afterthought. There
are good reasons for these omissions.  First, investigating the line
shifts requires knowing the systemic redshift of the quasar
accurately, whereas it is often the case that one simply takes the
peak of \civ\ as the redshift (thus {\em defining} the line shift to
be zero).  Second, while, like the BEff, we find that the blueshifts
are luminosity dependent (see Section~\ref{sec:c4prop}), the shifts
are seen at all luminosities, thus it might be forgiven if the
blueshift effect gets separated from the BEff.

In the context of the accretion disk-wind paradigm, it is as important to understand the range of wind parameters as it is to establish a connection to the underlying physical parameters (primarily mass and accretion rate).  We emphasize that the rapid growth (and general acceptance) of the disk-wind model \citep[e.g.,][]{mcgv95,elvis00,psk00} post-dates much of the work on the BELR (including \citealt{bg92}).  Indeed, \citet{mcgv95} and \citet{up95} were contemporaneous to within a month.  While many papers have explored the underlying {\em physical parameters} (e.g., $L/L_{Edd}$, $\dot{M}$, $M$; see previous references) there has been less work on understanding the actual {\em physics} in an accretion disk-wind scenario.  The exceptions to the rule \citep[e.g.,][]{Leighly04,clb06,lhj+07} focus on those objects whose SEDs have been studied in the most detail.  These also tend to be the most extreme objects within the AGN population --- which certainly are an important vector for further investigation, but the analysis still needs to be extended to the general population.

The fundamental premise herein is that the diversity in measurements
of BELR parameters is connected to the diversity of the disk-wind
parameters and one of our goals is to attempt to place the BEff,
\civ\ blueshifts, and the $L_{\rm UV}$--$\alpha_{\rm ox}$ into a
common picture of the accretion disk wind.  Indeed, it could be argued
that the luminosity-dependence of quasar SEDs {\em requires} the
rejection of a universal, static model for the BELR.  This statement
merely reflects the fact that the wind has a sensitive dependence to
both line-driving and ionizing photons that is expected to lead to
changes in the wind parameters as a function of the SED
\citep[e.g.,][]{psk00}.  As such, the debate between orientation
versus evolution presents a {\em false choice}: as quasars accrete more
mass and use up their available fuel, the SED (and thus the wind
parameters) will change.  The properties of quasars (the BELR in
particular) must evolve.  Yet there is no denying the axial symmetry
of quasars that leads to orientation effects.  A complete picture of
quasars (and their BELRs) requires an understanding of both of these
aspects.

This paper is structured as follows.  In Section~2, we describe the
data used in our investigations.  Section~\ref{sec:model} discusses
the physical context in which we evaluate our empirical results.
Section~\ref{sec:c4prop} summarizes the results for the \civ\ emission
line, while Section~\ref{sec:comp} considers these results in the form
of a series of composite spectra.  After a discussion of a number of
issues in Section~\ref{sec:discussion}, we present our conclusions in
Section~\ref{sec:unified}.

\section{Data}
\label{sec:data}

Our analysis is based on data from the 7th Data Release (DR7;
\citealt{aaa+09}) of the Sloan Digital Sky Survey (SDSS;
\citealt{yaa+00}).  The SDSS data were obtained with a dedicated
wide-field 2.5m telescope \citep{gsm+06} at Apache Point Observatory,
New Mexico with a 140-megapixel imaging camera \citep{gcr+98} and a
pair of fiber-fed multi-object double spectrographs.  The imaging was
carried out in five broad bands ($ugriz$; \citealt{fig+96},
\citealt{slb+02}) on photometric moonless nights of good seeing
\citep{hfs+01}.  The imaging data were then processed with a series of
pipelines \citep{lgi+01,pmh+03}, resulting in astrometric calibration
errors of $< 0.1''$ rms per coordinate, and photometric calibration to
better than 0.03 mag \citep{stk+02,ils+04,tkr+06}.  The photometry was
corrected both for Galactic extinction using the maps of \citet{sfd98}
and for spatial-dependent offsets using the \"{uber}-calibration
process described by \citet{psf+07}.  Our analysis relies on the point
spread function (PSF) asinh \citep{lgs+99} magnitudes.

Spectroscopic targets were selected from the imaging catalogs,
assigned to spectroscopic tiles \citep{blm+03}, and spectra were
obtained.  These data have been made publicly available in a series of
data releases culminating in DR7, which contains all of the data from
the SDSS-I and SDSS-II projects.  Main survey quasar candidates are
selected for spectroscopic followup as described in \citet{rfn+02}.
In addition to color selection, objects having 20\,cm radio
counterparts in the FIRST \citep{bwh95} survey are also targets (to
$i<19.1$).  Radio selection, while contributing less than $1$\% to the
color-selected sample \citep{imk+02}, is important for our
consideration of the differences between radio-loud and radio-quiet
quasars.

Our analysis specifically relies on the catalog of over 100,000
bona-fide quasars compiled by \citet{srh+10}.  These spectroscopically
confirmed quasars are restricted to those objects that meet a
traditional quasar definition ($M_i<-22$ measured in the rest frame,
and a full width at half maximum [FWHM] of lines from the broad line
region greater than $1000\,{\rm km\,s^{-1}}$).  A crucial aspect of
this investigation is the use of improved redshifts for SDSS quasars
as computed by \citet[hereafter HW10]{HW10}\footnote{HW10 published
  redshifts for the DR6 sample of quasars, whereas we are using the
  redshifts from an updated DR7 analysis (which is statistically
  identical to DR6).  The HW10 redshifts for the full DR7 quasar catalog can be obtained from http://www.sdss.org/dr7/products/value\_added/ .}.  These redshifts are determined by bootstrapping from (more) robust redshifts determined from the peak of the narrow component of [\ion{O}{3}] (to correct for the blueshift of [\ion{O}{3}] itself; \citealt{Boroson2005}) at lower redshifts and correcting for the same luminosity-dependent emission line effects that are the subject of this investigation.  While the corrections are small, for our investigation they are significant and, furthermore, allow us to determine the \civ\ blueshift for objects where \ion{Mg}{2} is redshifted out of the SDSS spectral window.
We note that the accuracy of
the redshifts used herein is unique to investigations involving the
\civ\ emission line as authors are often forced to use the \civ\ line
itself \citep[e.g.,][]{dhs+02,whd04} to determine the redshift of the
quasar (which also results in a small, but systematic bias in the
determination of luminosity).
One caveat is that we do {\em not} adopt the radio-loud specific redshifts that HW10 determine for radio-loud quasars --- for reasons that will become obvious in Section~\ref{sec:rlrq}.  
As \citet{srh+10} does not catalog all of the information about the emission lines that will be used in our investigation, we used the SQL interface to the Catalog Archive Server (CAS\footnote{http://cas.sdss.org/dr7/en/}) to extract this information for all of our objects from the {\tt SpecLine} table.  In the Appendix, we provide a simplified version of this query as an example in addition to giving the location of our input and output data files from our actual query.

We consider two samples of data.  The first largely follows the
primary sample analyzed by \citet{rvr+02}.  Specifically, we consider
SDSS-DR7 quasars with redshifts between 1.54 and 2.2 such that both
the \ion{C}{4} and \ion{Mg}{2} emission lines are well-measured in the
SDSS spectral range.  For these objects we can compare the redshift
determined from low-ionization \ion{Mg}{2} to the HW10 redshifts and
can confirm that the redshifts of HW10 are appropriate to use in
higher-redshift samples where \ion{Mg}{2} is unavailable.
As discussed in \citet{rvr+02}, we use the SDSS pipeline measurements for the emission lines and perform no further deblending or other re-reduction of the data.  The advantage of this approach is that all the spectra are treated in the same way using a well-established and robust algorithm.  A disadvantage is that further line deblending (such as treating the narrow line components of broad lines separately as suggested by \citealt{smd00}) is not possible.  However, the advantage of a large, uniformly processed sample outweighs this deficiency for the purpose of our statistical analyses.  We have confirmed that our results hold if we instead use the more sophisticated line deblending recently performed by \citet{Shen2010}.  The primary differences being that the SDSS pipeline allows only one Gaussian component to fit the \civ\ line (and another for each of the nearby \ion{He}{2} and \ion{O}{3}] lines), whereas \citet{Shen2010} allow up to three Gaussians to fit \civ\ alone.  The SDSS pipeline EQWs for \civ\ used herein are found to be systematically lower by 0.2 dex than those of \citet{Shen2010}; this offset is not dependent on the other parameters considered herein and has no affect on our results.

To minimize outliers left by the automated processing in terms of the
measurement uncertainties ($\sigma$), we apply the following
restrictions to the sample: ${\rm FWHM}_{\rm CIV} > 1000\;{\rm
  km\,s^{-1}}$ and ${\rm FWHM}_{\rm MgII} > 1000\;{\rm km\,s^{-1}}$ to
consider only bona-fide broad line objects; $\sigma_{\lambda_{\rm
    CIV}} < 10\;{\rm \AA}$ and $\sigma_{\lambda_{\rm MgII}} < 15\;{\rm
  \AA}$ to eliminate objects with questionable line centers; and ${\rm
  FWHM}_{\rm CIV} > 2\sigma_{{\rm FWHM}_{\rm CIV}}$, ${\rm EQW}_{\rm
  CIV} > 2\sigma_{{\rm EQW}_{\rm CIV}}$, ${\rm EQW}_{\rm CIV}>5\;{\rm
  \AA}$, ${\rm FWHM}_{\rm MgII} > 2\sigma_{{\rm FWHM}_{\rm MgII}}$,
${\rm EQW}_{\rm MgII} > 2\sigma_{{\rm EQW}_{\rm MgII}}$ to eliminate
objects with lines detected with questionable significance.  
We also restrict the error on the \ion{C}{4} blueshift to be less than
$600\;{\rm km\,s^{-1}}$ (as computed from the pipeline-determined
errors in the central wavelengths of \ion{C}{4} and \ion{Mg}{2}); the
mean error is $180\;{\rm km\,s^{-1}}$.  These restrictions are
intended to cull errors in the pipeline fitting process (e.g., due to
glitches in the spectra) without overly biasing the sample against
intrinsically weak lines.  We make no explicit cut on the signal-to-noise ratio (S/N) of the
spectra, but will show in Section~\ref{sec:c4prop} that our results
are unaffected by our inclusion of low S/N spectra.  Finally, we have
removed all known broad absorption line quasars as cataloged by
\citet{Allen10}.  While this BALQSO catalog was constructed for the
DR6 sample, it removes $\sim$85\% of the BALQSOs in the DR7
spectroscopic footprint.  In all, we are left with 15,779 SDSS quasars
with well-measured \ion{C}{4} and \ion{Mg}{2} emission lines; we will
refer to this sample as ``Sample A''.  These quasars were drawn from
31,978 quasars with both \civ\ and \mgii\ coverage; most of the
reduction comes from our limits on the error in the \ion{C}{4}
blueshift.  By comparison, our \ion{C}{4}--\ion{Mg}{2} sample in
\citet{rvr+02} had only 794 quasars.

As our primary goal is to investigate quasars as a function of
\ion{C}{4} emission line properties (specifically the blueshift and
the equivalent width), the requirement of a well-measured \ion{Mg}{2}
line is overly restrictive.  It is necessary simply to enable a
comparison to the systemic redshift (or something close to it).  Since
HW10 have developed a process for correcting systematic errors in the
redshifts of SDSS quasars (at the level of a few hundred ${\rm
  km\,s^{-1}}$), it is possible to reference our \ion{C}{4}
measurements to those redshifts instead.  Doing so allows us to loosen
some of the above restrictions and extend our sample to higher
redshifts for a total of 35,391 quasars; we will refer to this sample
as ``Sample B''.  Sample B allows us to explore the luminosity
dependences of emission line properties over a larger dynamic range.
However, it comes with the trade-off of being more sensitive to the
$L$--$z$ degeneracy inherent to a flux limited survey (indeed there
are 2 distinct flux limits in the SDSS quasar selection algorithm),
and to redshift- and color-dependent selection effects at $z>2.2$
where the SDSS sensitivity exhibits strong features \citep{rfn+02,
  rsf+06} --- particularly with respect to differences in the
selection of radio-detected quasars at high redshift.  As such, we
will utilize both samples according to which is more appropriate for a
particular analysis.

\begin{figure}[h]
\epsscale{1.0}
\plotone{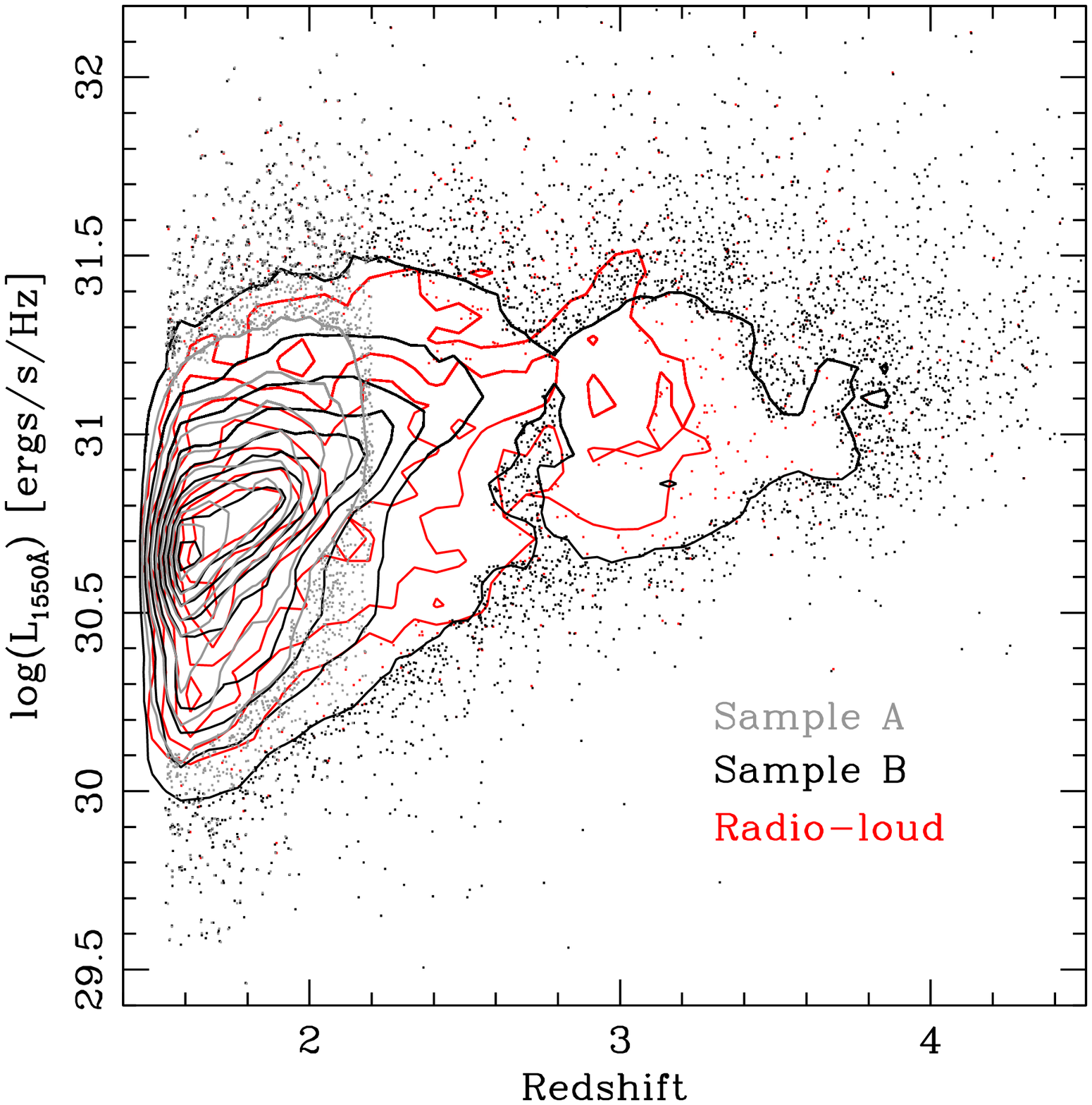}
\caption{Luminosity and redshift distributions of Samples A ({\em grey}) and B ({\em black}).  The luminosity is the continuum luminosity at $1550\,{\rm \AA}$ as determined from the SDSS emission line fitting procedure.  The redshifts are from HW10.  Sample A is more restrictive and less subject to bias.  Sample B extends to higher redshift and includes more high luminosity objects, but can suffer from certain biases (e.g., a larger radio-loud fraction at $z\sim2.5$, where SDSS color selection is very inefficient).  The red points/contours indicate the radio-loud quasars in the sample.}
\label{fig:LzHW}
\end{figure}

Figure~\ref{fig:LzHW} shows the luminosity (in the form of a continuum luminosity at $1550\,{\rm \AA}$) and redshift distribution of both Samples A and B.  Since the distinction between radio-quiet (RQ) and radio-loud (RL) quasars will be a common theme throughout this paper, we will distinguish between these two populations from the outset.  In general, we will classify quasars as radio-loud if they have peak radio luminosities that exceed $\log L_{\rm 20\,cm} = 32.5\; {\rm ergs\,s^{-1}\,cm^{-2}\,Hz}$ \citep[e.g.,][]{gkm+99}, where we have assumed a radio spectral index of $\alpha_{\nu}=-0.5$.  With this definition, the FIRST survey \citep{bwh95} is complete to radio-loud quasars to $z\leq2.7$; at higher redshifts, FIRST may not be deep enough to detect all radio-loud quasars.  In terms of defining quasars as radio-loud by their optical to radio flux ratio \citep[e.g.,][]{kss+89}, our combined SDSS+FIRST sample is complete to $i=18.9$ (where a typical radio-loudness cut is $\log({\rm optical}/{\rm radio})\ge1$, the FIRST flux limit is $1\,{\rm mJy}$, and $AB=18.9$ corresponds to $10\,{\rm mJy}$).  Fainter than this it is possible for bona-fide radio-loud objects to be fainter than the FIRST flux density  limit, thus some caution is needed in the treatment of radio-loud and radio-quiet quasars.  As such, when it is important that we know that objects are indeed radio-quiet (and not simply undetected at the faint limit of FIRST) we will restrict the samples to $i=18.9$.  We further restrict the radio-quiet sample to $\log L_{\rm 20\,cm} < 32.0\; {\rm ergs\,s^{-1}\,cm^{-2}\,Hz}$, leaving a small gap of radio-intermediate sources.  Figure~\ref{fig:LzHW} shows that adopting these restrictions produces a radio-loud sub-sample that has no large scale biases with respect to quasar selection.

\section{Model Presumptions}
\label{sec:model}

While our analysis is not specifically model dependent, it is
important to understand the framework in which we consider our
results.  Our assumption is that of a 2-component model of the BELR,
specifically a ``wind'' and a ``disk'' component.  These terms are used to group like features in the BELR --- regardless of the disk and wind interpretations; however, they are indicative of our assumptions about the location of the dominant contribution to the BELR in each of these classes of objects.  For the wind, we adopt a
Murray \& Chiang style rotating accretion disk wind that is being
accelerated outwards by radiation line driving; see
\citet{mcgv95}. For a general introduction to AGN outflows see
\citet{bp82}, \citet{psk00}, \citet{everett05} and references therein.
In such a model the wind component of the high-ionization emission
lines is single-peaked due to radiative transfer effects.  The
structure of the wind is balanced by the trade-off between UV and
X-ray photons, where UV photons accelerate the gas due to radiation
line pressure through UV line transitions
and the X-ray photons can strip the gas of electrons, thereby reducing
the line driving.  As such, the detailed shape of the SED is a crucial
aspect of the model.  Except for Section~\ref{sec:lme}, we do not
consider in detail what determines the SED (e.g., mass, accretion
rate, etc.), but rather concentrate on how the shape of the SED
affects the nature of the wind.

We note that \citet{gaskell09} has recently commented that outflows are ``ruled out'' 
by early reverberation mapping results.  Specifically, it has been
shown\footnote{For NGC~5548, which has both an atypical continuum and
  a very large \civ\ EQW.} that the red wing of the \civ\ line
leads the blue wing \citep{kab+95}, whereas the opposite might be
expected in a pure outflow scenario.  While we agree that this is the
case for a pure outflow, as emphasized by \citet{szm+00}, this is not
an issue for a disk-wind model (which is not a pure outflow).  Indeed
it has been shown that such a wind can explain the reverberation
mapping results \citep{cm96} in addition to providing an explanation
for the single-peaked nature of the \civ\ line.

For the disk component
we adopt a ``filtered continuum'' model \citep{Leighly04,lc07}, in which the disk component may not see the same continuum that we see (or that the wind component sees) as the SED is first transmitted through the wind before reaching the disk.  Transmission through the wind absorbs part of the SED, making it effectively softer (e.g., Fig.\ 9 in Leighly 2004).  This assumption is important as it allows a coupling between these otherwise distinct regions of the BELR.  

The general idea of our two-component model is similar to that of \citet{Collin06} and one possible manifestation is illustrated in Figure~15 of \citet{Leighly04}, except that here we assume that there is always a wind component (which simply may not always be dominated by radiation line driving) and we make no assumptions about the geometry of the X-ray source or the structure of the inner disk.  Some recent investigations have presented evidence for an infalling component of the BELR \citep{hwh+08a,hwh+08b,fhw+09}. In general we do not consider such a component as shedding the angular momentum of the accreting gas is already a problem even without postulating an infalling component; however, the evidence for such a component is certainly of interest and should be investigated further.

In our adopted picture, the range of observed BELR parameters reflects the
competition between radiation line driving and winds driven by other
mechanisms (e.g., magneto hydrodynamic [MHD] driving); see \citet{Proga03}.
Since the disk component of the BELR is filtered through the wind, in our framework changes in the wind are also a source of the (generally anti-correlated) changes in the disk component.  In the context of this model, the EQW of the \civ\ emission line is a diagnostic of the total amount of line emitting gas and ionizing radiation in the BELR and the blueshift of \civ\ provides a measure of the relative strength of the wind and disk components.  Given this interpretation, we do not consider trends with \civ\ FWHM as it is not clear what physical meaning the FWHM would have in this two-component picture.

\section{Characterization of \ion{C}{4} Properties}
\label{sec:c4prop}

\subsection{Distribution of \ion{C}{4} EQW and Blueshift}

\begin{figure}[h]
\epsscale{1.0}
\plotone{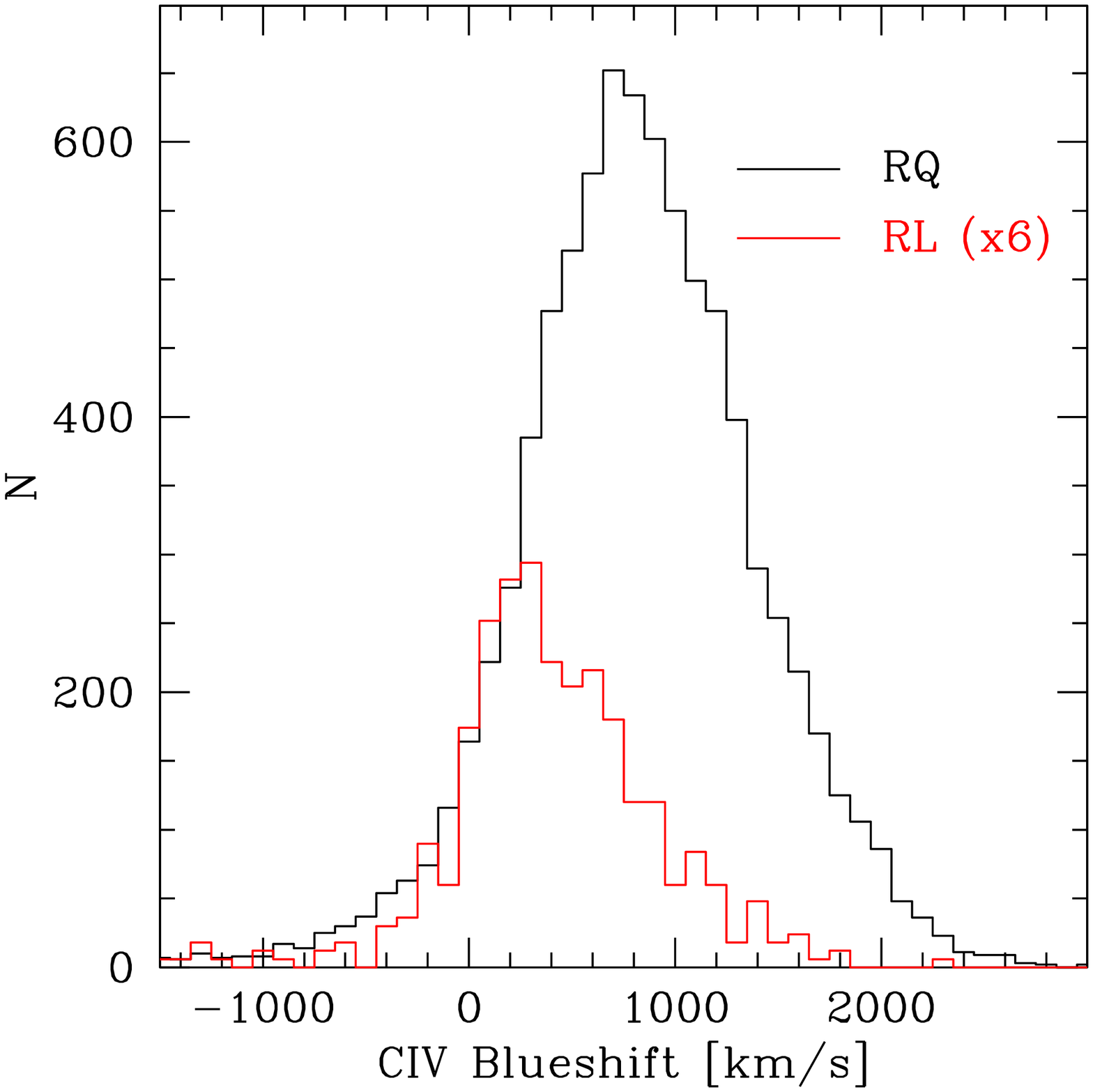}
\caption{Distribution of \ion{C}{4} blueshifts for radio-quiet ({\em black}) and radio-loud ({\em red}) quasars for Sample A, further restricted to $i<18.9$. Scaling the radio-loud sample by a factor of $\sim6$ causes the low-velocity edges of the RQ and RL samples to roughly match.  Note that RL quasars make up about 5--20\% of the total quasar population \citep{jfi07} depending on redshift and luminosity.  While there are objects with statistically significant negative blueshifts (7\% for the RQ objects and 15\% for the RL objects), in general, the \ion{C}{4} blueshift is ubiquitous.}
\label{fig:c4bhist}
\end{figure}

The two key diagnostics that we will use in our analysis are the
blueshift and equivalent width (EQW) of the \ion{C}{4} emission line.
As such, we begin by examining the distributions of these two
properties.  
Figure~\ref{fig:c4bhist} shows the distribution of \ion{C}{4} blueshifts for the more restrictive Sample A (further limited to objects with $i<18.9$ to avoid the radio-loud incompleteness discussed above).  Note that the sign convention we use for blueshifts is the same as \citet{rvr+02}, but is different than some other authors use.  While the term ``blueshift'' is derived from the fact that there appears to be a component of the BELR moving towards us and the line peak is shifted bluewards in wavelength, fundamentally this is an outflow from the quasar.  Thus we adopt the natural quasar frame, instead of our own artificial frame, and assign positive velocities (i.e., outflows in the quasar frame) to objects with large \ion{C}{4} blueshifts.

In Figure~\ref{fig:c4bhist} a striking aspect of the \civ\ blueshift distribution is that it appears to be nearly ubiquitous \citep[e.g.,][]{szm+00}.  Only 7\% of RQ objects are negative outliers, while 15\% of RL objects are negative outliers\footnote{Measured blueshift $\le0$ {\em and} more than $3\sigma$ from the mean.} and even those negative values may simply be due remaining systematics in the redshift corrections of HW10.
The median blueshift for radio-quiet quasars is $814\;{\rm km\,s^{-1}}$ and, while radio-loud quasars show much smaller blueshift,
even their median is $364\;{\rm km\,s^{-1}}$.  We have scaled the RL sample by a factor of 6 in order to align the left-hand sides of the RL and RQ histograms for the sake of comparing the two distributions.  As the RL population is as low as $\sim5\%$ and as high as $\sim20\%$ of the total depending on redshift and luminosity \citep{jfi07}, other scaling factors could be more appropriate.  It is often argued that these blueshifts are an indication of an outflowing/wind component in AGNs \citep[e.g.,][]{gas82,Leighly04}.  If so, we must conclude that much of the \civ\ emission is produced in the wind and that essentially all quasars have a significant wind component to the BELR.  In that case, the fact that RL and RQ quasars have different distributions suggests that they have, on average, different wind properties; see Section~\ref{sec:rlrq} for further discussion of this issue.

\begin{figure}[h]
\epsscale{1.0}
\plotone{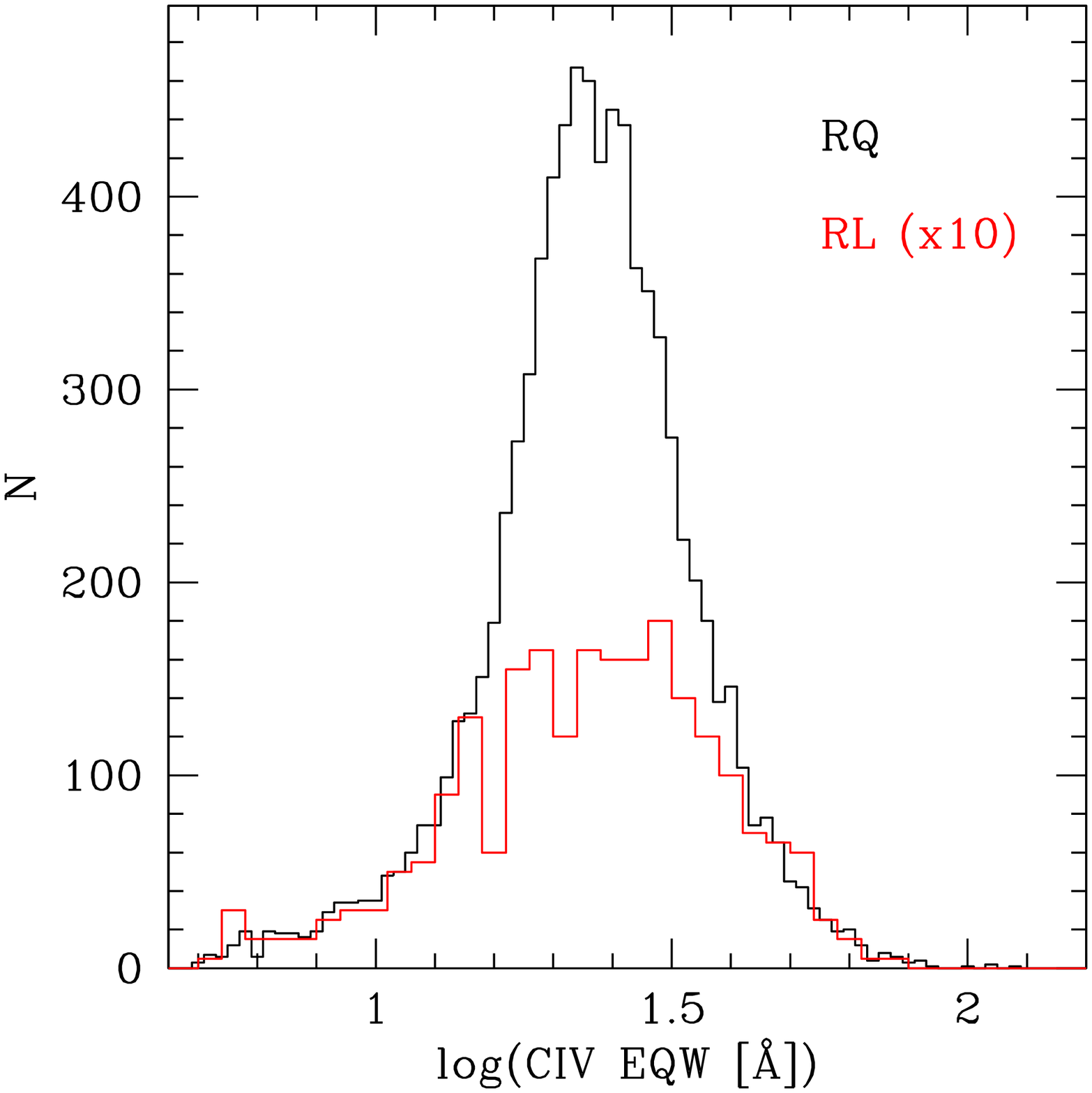}
\caption{Distribution of \ion{C}{4} equivalent widths for radio-quiet ({\em black}) and radio-loud ({\em red}) quasars for Sample A (with $i<18.9$).   Here the RL sample has been scaled by a factor of 10 to highlight the differences in the central part of the distribution once the wings are aligned.}
\label{fig:c4eqw}
\end{figure}

\begin{figure}[h]
\epsscale{1.0}
\plotone{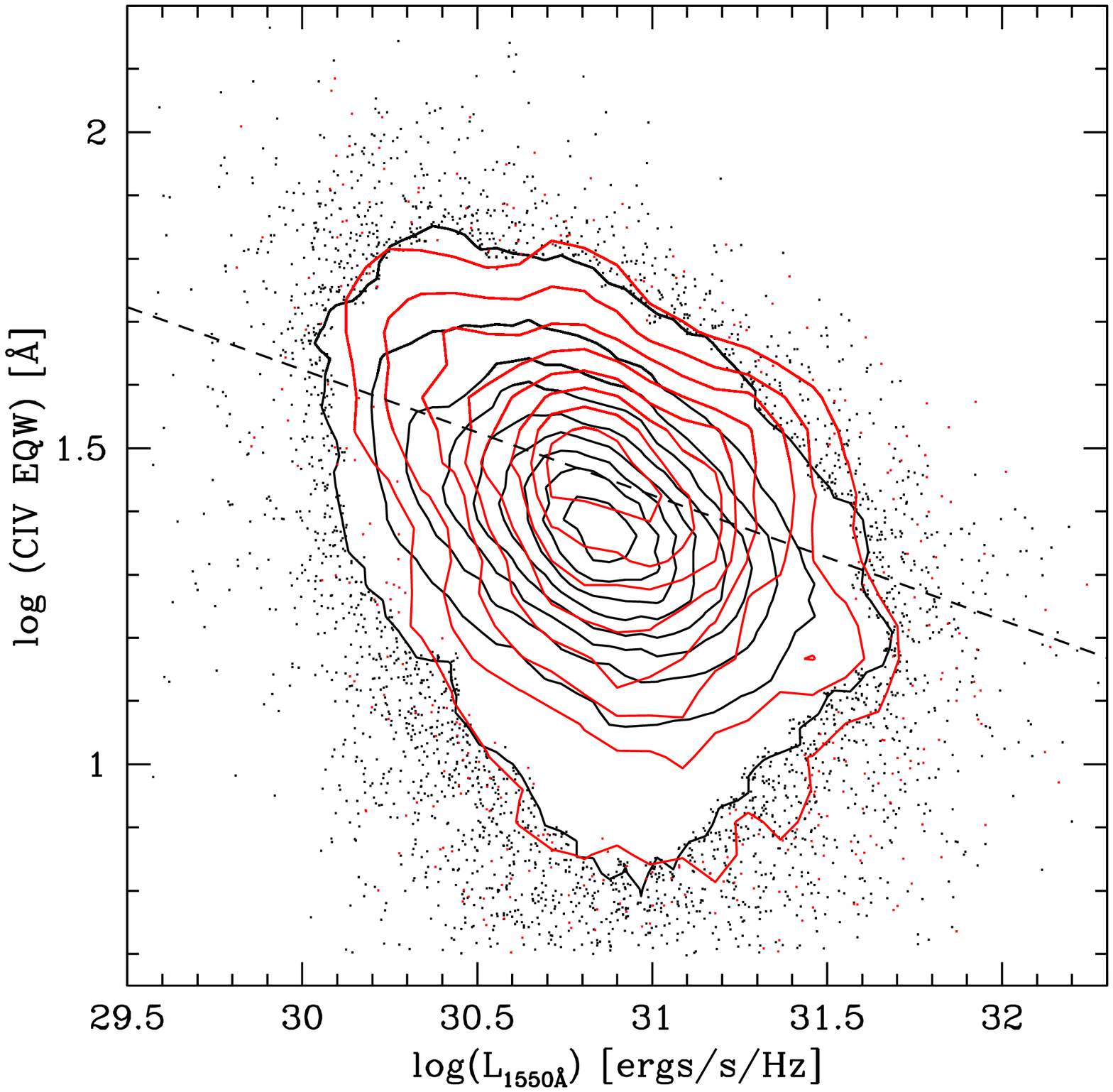}
\caption{Decrease of \ion{C}{4} EQW with luminosity, generally known as the Baldwin Effect.  Both RL ({\em red}) and RQ ({\em black}) samples show this trend.  Here we use Sample B, which  allows a larger range of luminosities to be probed.  The dashed line shows the best fit from \citet{wvb+09}, which spans a range in luminosity of $27.81< \log L_{\nu} (2500{\rm \AA})<33.04$.  Note that the \citet{wvb+09} line has been lowered by 0.2 dex to account for the differences in EQW values as measured by the SDSS spectroscopic pipeline \citep{Shen2010}.
}
\label{fig:beff1550}
\end{figure}

Figure~\ref{fig:c4eqw} shows the distribution of \ion{C}{4} equivalent widths.  Unlike the \ion{C}{4} blueshifts, both the RQ and RL quasars span the full distribution of measured values.  This is of interest considering that the BEff is a trend of decreasing \ion{C}{4} equivalent width with luminosity.  Since both the RQ and RL populations span a large range of equivalent widths, we might expect both populations to exhibit a Baldwin Effect.  Indeed, the original BEff work was dominated by flat-spectrum RL quasars (and 60\% of the sample being radio-loud)\footnote{Investigations of \civ\ that are dominated by objects in the {\em HST} archive and/or by ``PG''  quasars \citep{bg92} are biased towards RL quasars by up to a factor of 5 \citep{sbm+07} as compared to our sample.} and a BEff for both RL and RQ quasars has already been seen.  As such, in Figure~\ref{fig:beff1550} we plot the EQW of \ion{C}{4} versus luminosity, specifically, $L_{\nu} (1550\,{\rm \AA})$.
While the dynamic range of our SDSS sample is relatively small, 
and both variability \citep{wvb+09} and our use of SDSS pipeline measurements for the \ion{C}{4} EQW may cause some scatter, the general trend of the BEff is clear.  This is true for both RL and RQ quasars.  We emphasize that the choice of luminosity is important to the BEff \citep[e.g.,][]{Green96}.  Here our choice of $L_{\nu}(1550)$ is merely a matter of convenience as it is measured (uniformly) for all of the objects in our sample.  

\begin{figure}[h]
\epsscale{1.0}
\plotone{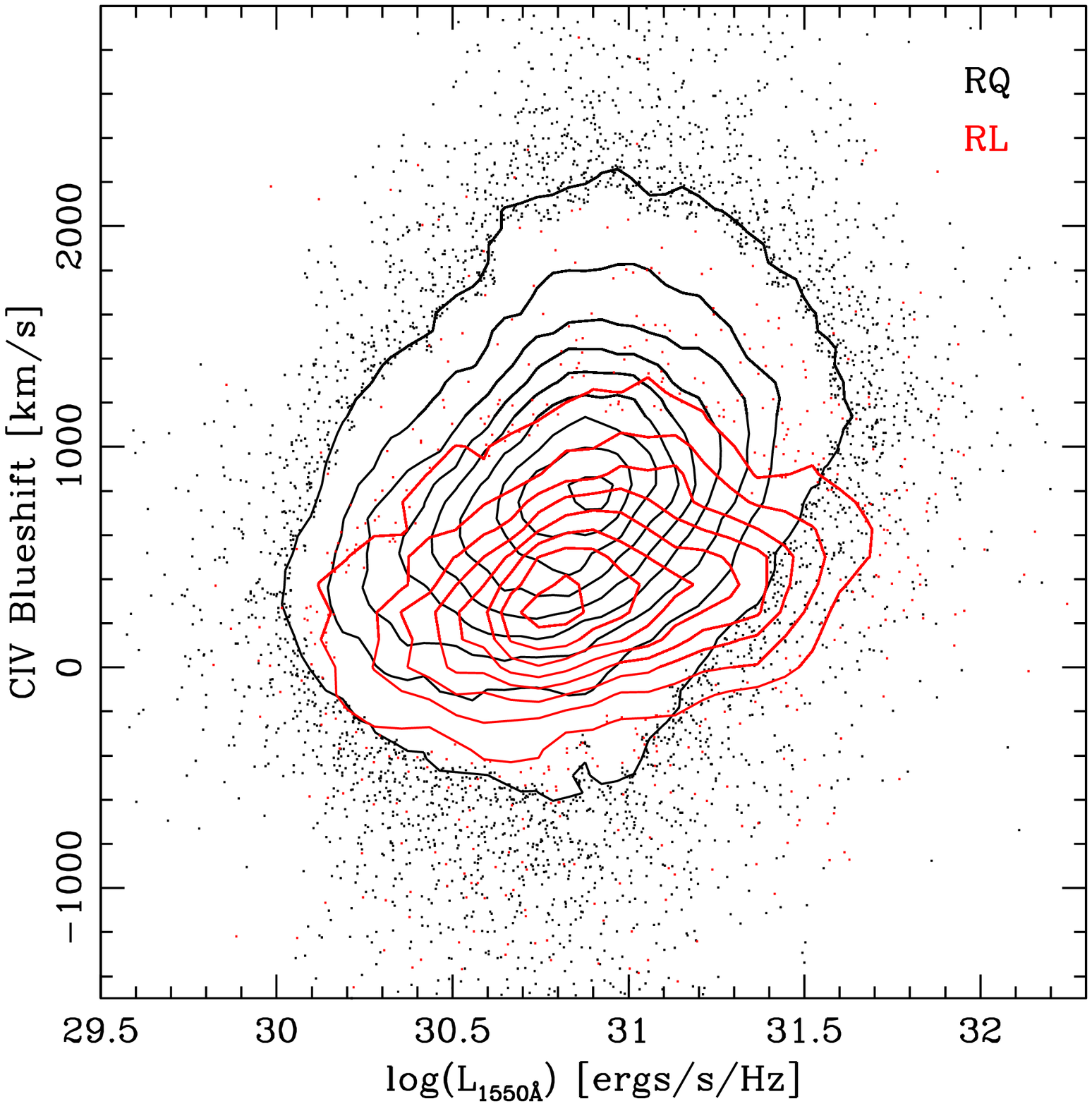}
\caption{\civ\ blueshift vs.\ luminosity for RQ quasars ({\em black}) and RL quasars ({\em red}).  While the trend with luminosity for RL quasars is shallower, both RQ and RL quasars have larger blueshifts at higher luminosities.}
\label{fig:c4b1550}
\end{figure}

Figure~\ref{fig:c4b1550} shows the same range in luminosity as for Figure~\ref{fig:beff1550}, but this time against the \civ\ blueshift.  Again we see that both RL and RQ quasars participate in the trend, although the RL trend has a shallower slope.  As with the BEff, the scatter is large, but, generally speaking, luminous quasars are more likely to have larger \civ\ blueshifts.  We note that the $L-z$ distributions (Fig.~\ref{fig:LzHW}) of objects with extrema in their \civ\ properties (high EQW or large blueshift) are consistent with their luminosity distributions shown in Figures~\ref{fig:beff1550} and \ref{fig:c4b1550} and the $L-z$ degeneracy inherent to this flux limited sample.  As such, there is no reason to believe that trends with luminosity are instead trends with redshift (see Section~\ref{sec:xray}).

\begin{figure}[h]
\epsscale{1.0}
\plottwo{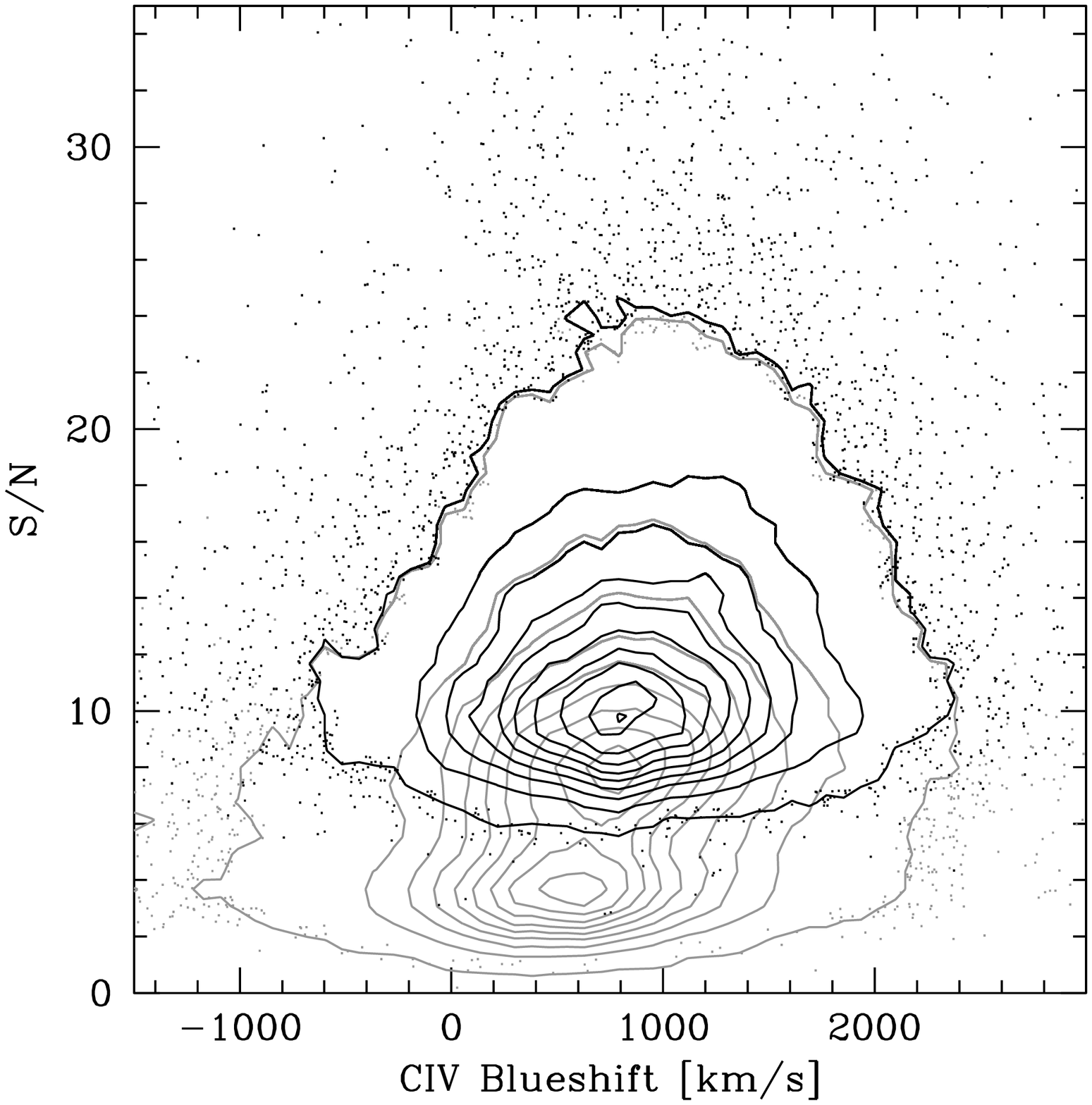}{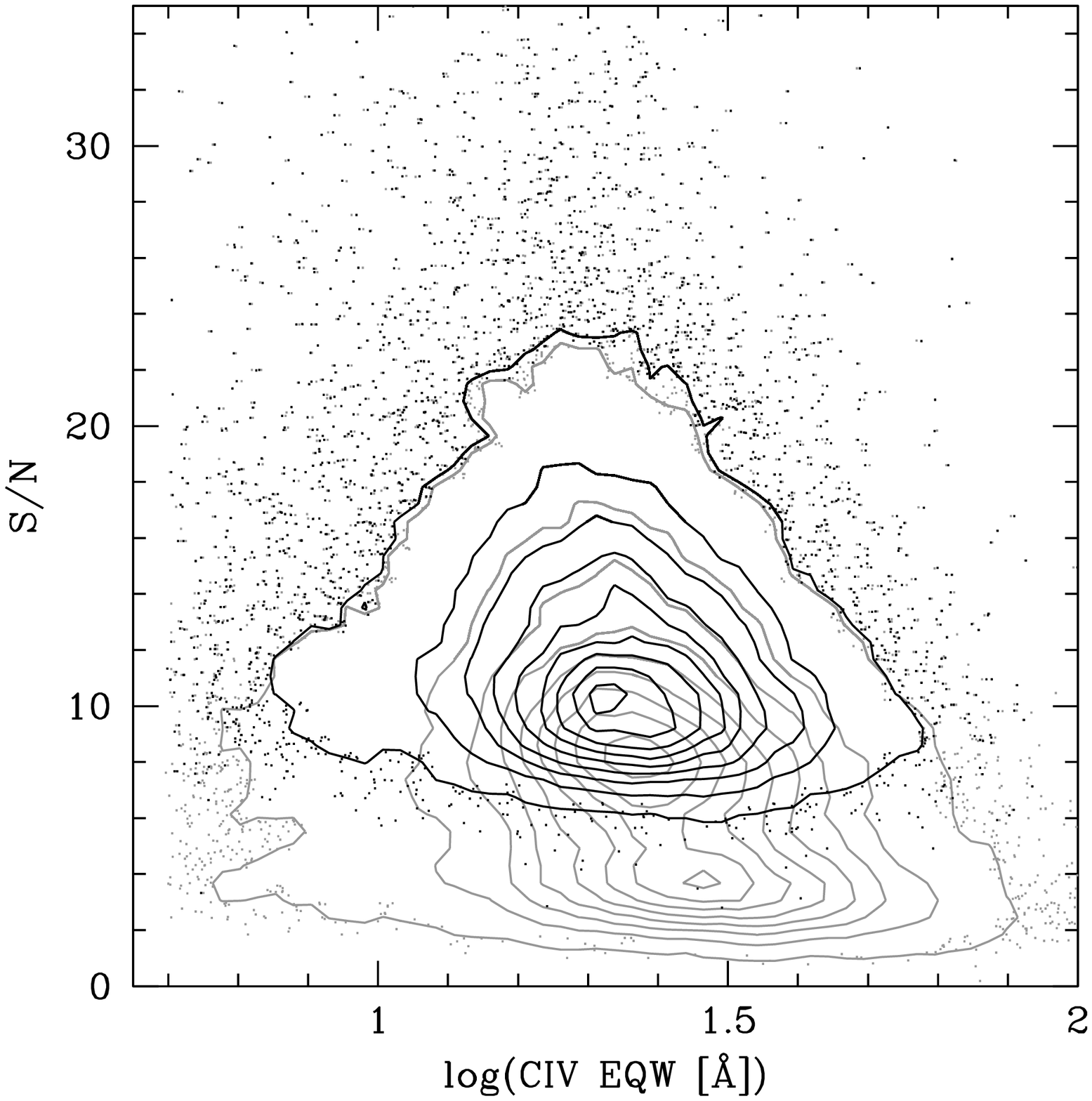} 
\caption{({\em Left:}) \civ\ blueshift vs.\ spectral S/N (synthesized $i$-band).  ({\em Right:}) \civ\ EQW vs.\ spectral S/N.  In both panels, gray contours/points indicate all of the objects that meet the cuts described in Section~\ref{sec:data}.  Black contours/points are also restricted to objects with $i<18.9$ where the RL sample is complete; these are the samples shown in Figures~\ref{fig:c4bhist} and \ref{fig:c4eqw}. The increase of the mean \civ\ blueshift and decrease of the mean \civ\ EQW with S/N are seen even for the brighter objects, which indicates that these changes are due to the observed luminosity dependence of these parameters and are not a S/N-dependent measurement bias.}
\label{fig:sn}
\end{figure}

As the SDSS quasar sample is dominated by low S/N spectra near the flux limit of the spectroscopic survey, it is important to establish that our results are not biased by automated measurements of line parameters in low S/N data.  As such, in Figure~\ref{fig:sn} we show the S/N of the synthesized $i$ band from the SDSS-spectra versus both the \civ\ blueshift and EQW measurements.  We find no unexpected trends with S/N, indicating that our results are not affected by biases in the spectral processing as a function of S/N (e.g., the high/low blueshift objects are not all at low S/N).  

\begin{figure}[h]
\epsscale{1.0}
\plotone{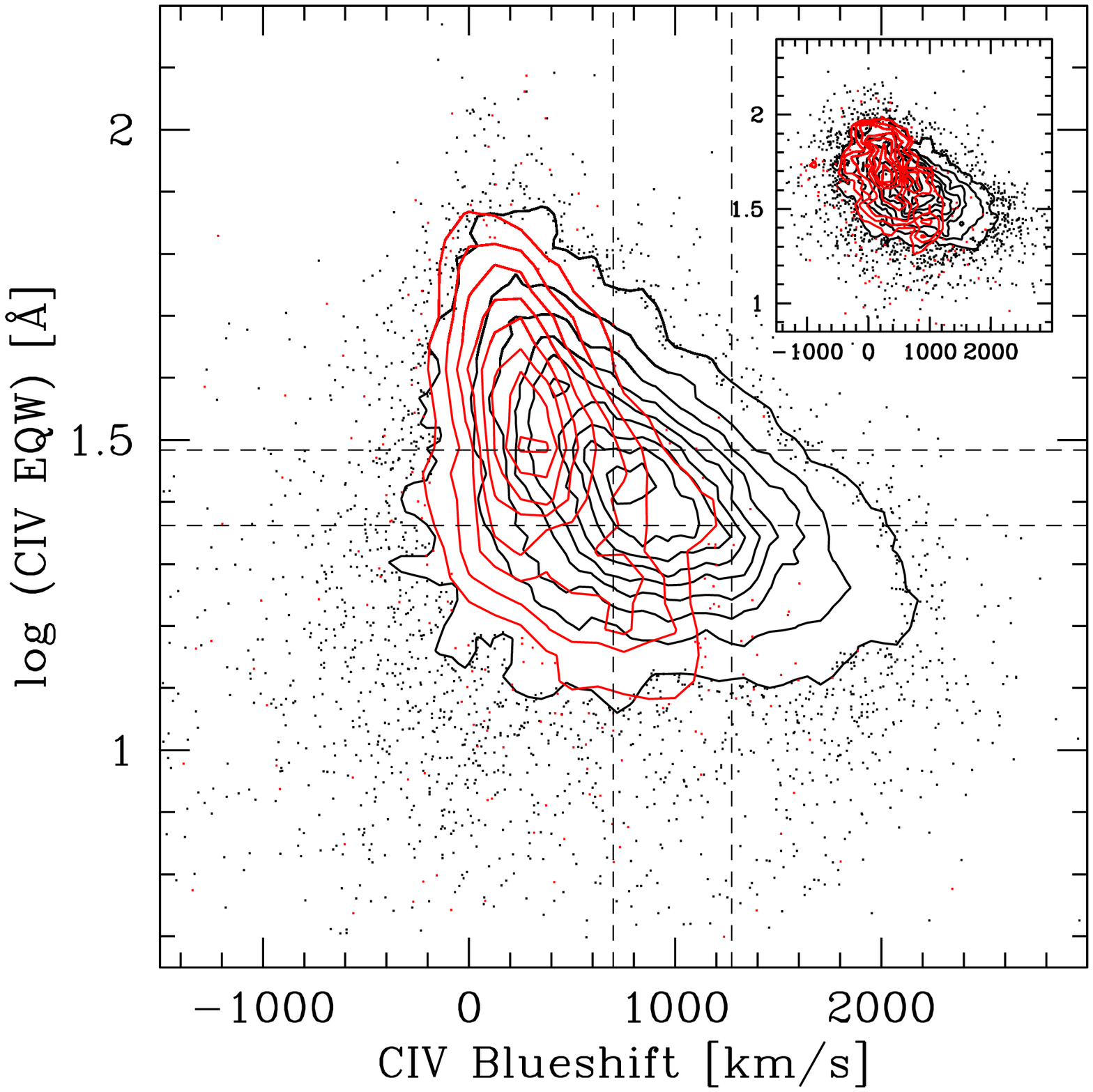}
\caption{\civ\ EQW vs.\ \civ\ blueshift.  RQ quasars are shown as black contours/dots, while RL quasars are shown as red contours/dots.  The dashed lines indicate the divisions used to create composite spectra in Section~\ref{sec:comp}.  Apparently there are no quasars with large blueshift and strong \civ\ emission lines; this is highly unlikely to be a selection effect.  The distributions remain unchanged when the samples are restricted to objects with high S/N.  The insert in the upper right shows the same plot, but instead using the data from \cite{Shen2010}.  The only differences are the aforementioned offset of the EQW distribution by 0.2 dex and a restriction to S/N$>10$ objects for the sake of clarity.  The basic properties of the distributions are unchanged.}
\label{fig:c4beqw}
\end{figure}

The similarity between the trends with luminosity in Figures~\ref{fig:beff1550} and \ref{fig:c4b1550} begs the question of relationship between the \civ\ blueshift and the BEff.  For example, is the reduction in EQW biased towards the red wing of the line such that these measurement are correlated (perhaps even having the same origin; \citealt{rvr+02,Richards2006})?    \citet{msd+96} found a clear trend for RQ quasars, but not for RL quasars (as might be expected given our finding that RL quasars span only a small range in blueshift).   Figure~\ref{fig:c4beqw} shows that large blueshift quasars do indeed have weak \civ\ and small blueshift quasars have strong \civ, but the trend is clearly not one-to-one as there are also quasars with small blueshifts {\em and} weak emission lines.  The distribution is unchanged even if we restrict the sample to those spectra with S/N$>10$.  Thus the blueshifting of \civ\ and the BEff are apparently not of the same origin.
It is true, however, that strong lines and large blueshifts are
mutually exclusive, leaving the distribution with a characteristic
``missing wedge''.  Curiously the RL population tends toward a
specific corner of the RQ parameter space rather than being a distinct
distribution as has also been noted by \citet{szm+00}.  In addition,
it seems that the RL quasars follow the same trend towards larger
blueshifts and weaker lines with increasing luminosity as do the RQ
quasars.  We will discuss the RL distribution in more detail in
Section~\ref{sec:rlrq}.  Note that, in Figure~\ref{fig:c4beqw} the UV
luminosity increases both downwards and to the right; see
Figures~\ref{fig:beff1550} and \ref{fig:c4b1550}.

\subsection{X-ray Properties}
\label{sec:xray}

\begin{figure}
\epsscale{1.0}
\plotone{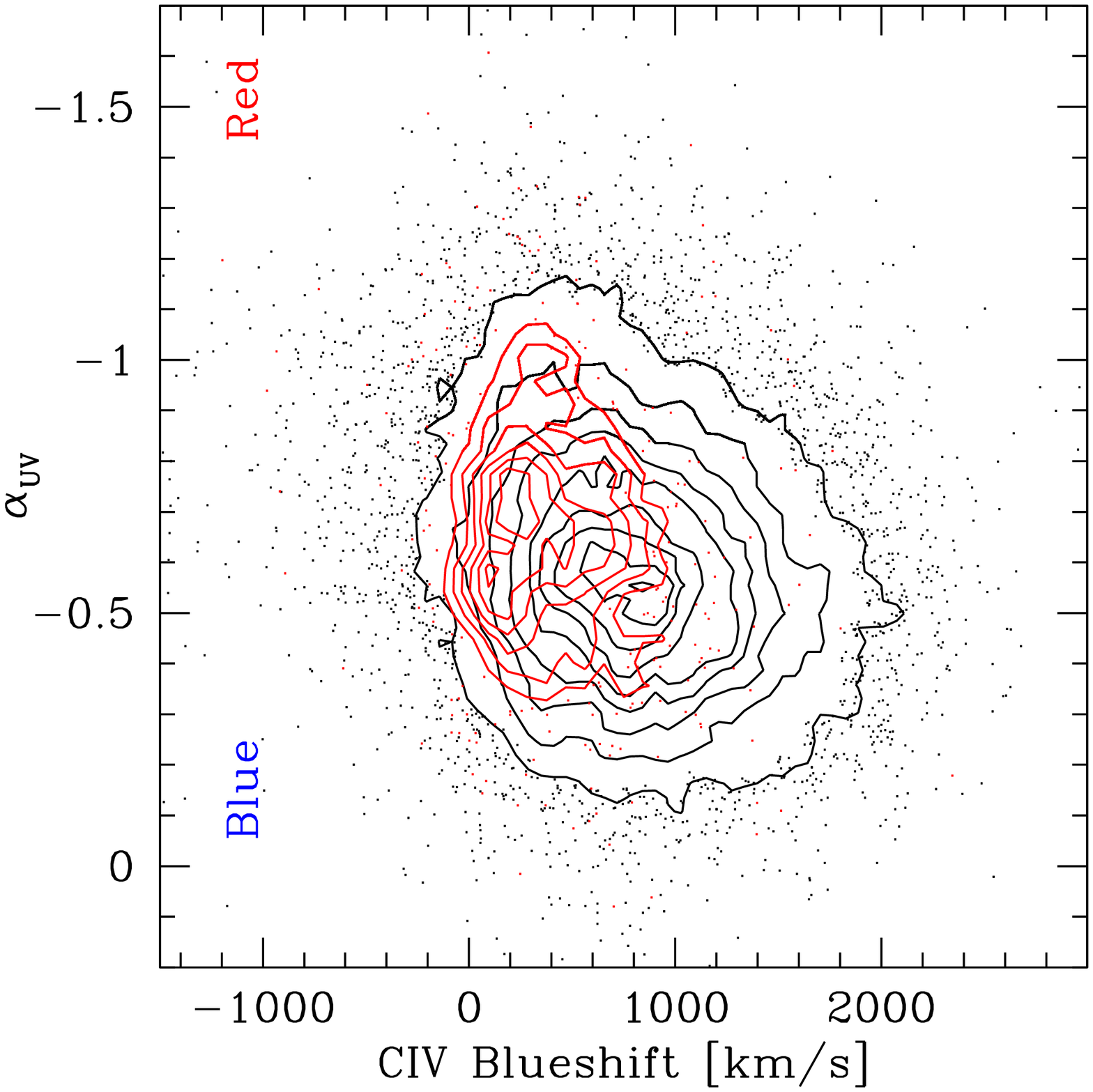}
\caption{\civ\ blueshift vs. $\alpha_{\nu}$ for Sample A.  RQ quasars are shown in black.  RL quasars ({\em red}) tend not only towards smaller \civ\ blueshifts as seen above, but also to somewhat redder colors.} 
\label{fig:c4balphanu}
\end{figure}

In \citet{grh+05} we presented another distribution with a similar missing wedge, specifically the color-blueshift distribution, where color is measured either by the spectral index, $\alpha_{\rm uv}$, or determined from the redshift-corrected broad-band colors, $\Delta (g-i)$ \citep{rhv+03}.  We showed that large blueshift quasars are less likely to have red continua (whether intrinsic or dust reddened).  In Figure~\ref{fig:c4balphanu}  we demonstrate this effect using Sample A.  Here we also find that RL quasars not only have smaller \civ\ blueshifts on average, they also tend to be slightly redder than the average RQ quasars (but still within the distribution of RQ quasars as a whole).  \citet{grh+05} also found that the X-ray spectral index, $\Gamma (\equiv 1-\alpha_x$), is correlated with $\alpha_{\nu}$ such that bluer quasars tend to have softer X-ray spectra at 0.5--8\,keV (see also \citealt{gkl+10}).  This finding may be important in the overall context of the BG92 eigenvector parameter space as $\Gamma$ (measured at 0.2--2\,keV) is one of the known Eigenvector 1 (EV1) correlates \citep{lfe+97}.

\begin{figure}
\epsscale{1.0}
\plottwo{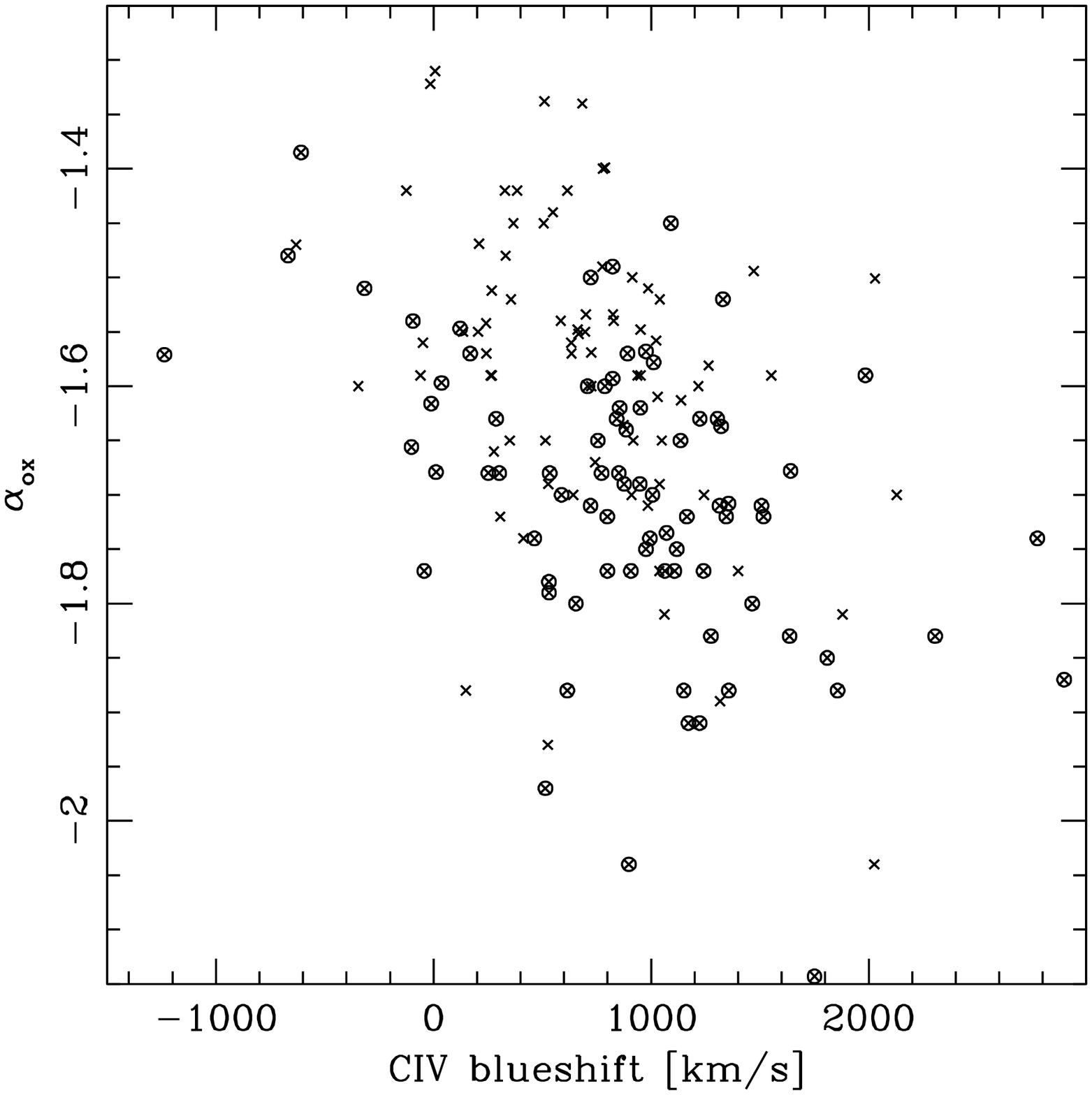}{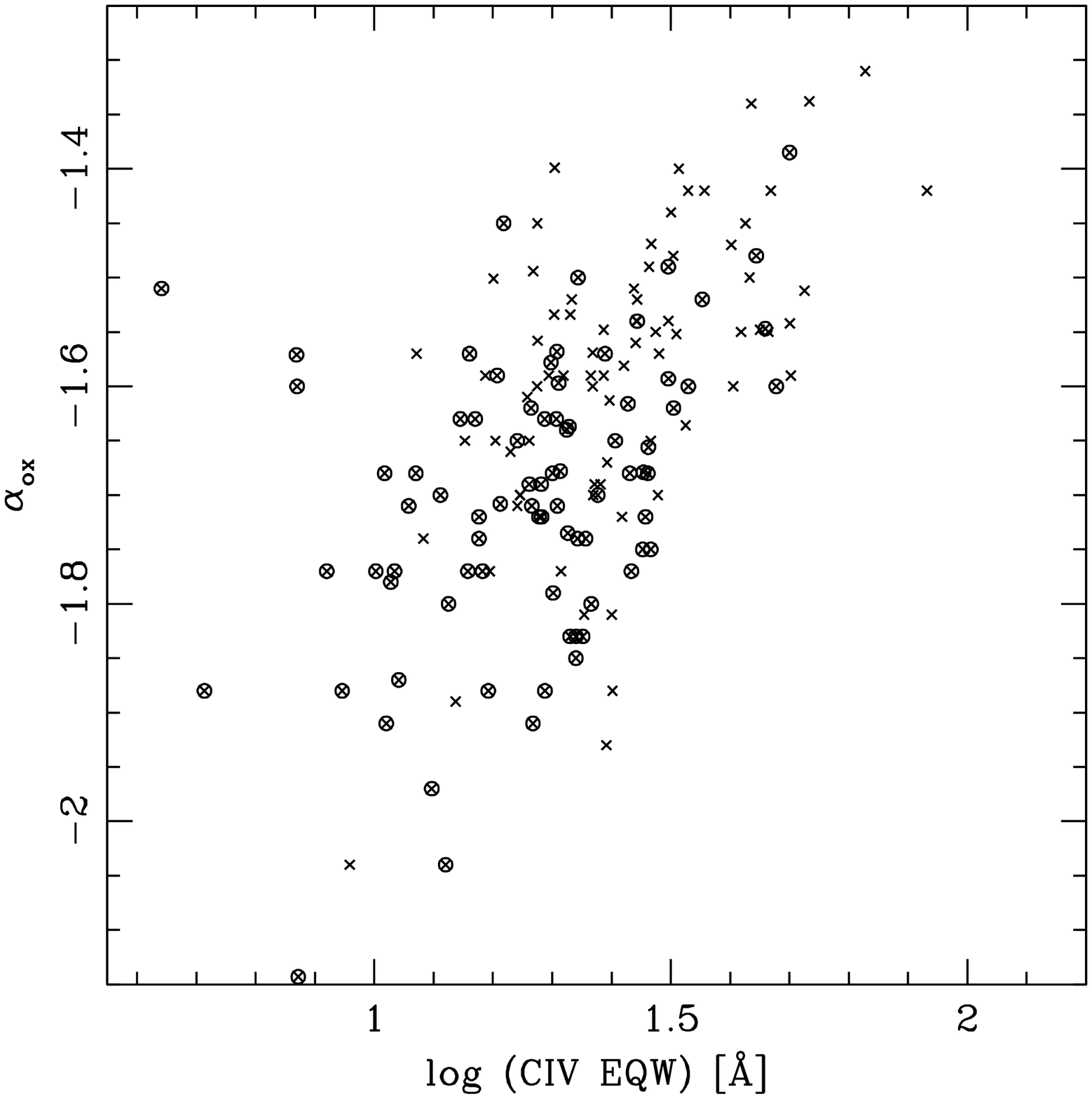}
\caption{$\alpha_{\rm ox}$ vs.\ \civ\ blueshift ({\em Left}) and \civ\ EQW ({\em Right}). Object with more negative values of $\alpha_{\rm ox}$ (taken from \citealt{wvb+09}) are apparently weaker in X-rays (in a relative sense) and have weaker \civ\ lines that are more strongly blueshifted.  Circled crosses indicate those spectra with S/N$>10$.}
\label{fig:wu09}
\end{figure}

In addition to $\Gamma$ it is interesting to consider the \civ\ parameters as a function of the optical-to-X-ray flux ratio, $\alpha_{\rm ox}$.  
A full X-ray investigation is beyond the scope of this paper, but we can use the sample compiled by \citet{wvb+09} to show some trends.  In Figure~\ref{fig:wu09} we show how $\alpha_{\rm ox}$ correlates with both \civ\ blueshift and \civ\ EQW: quasars that are relatively weaker in the X-ray (at 2\,keV) have larger \civ\ blueshifts and weaker \civ\ EQWs.  The correlation between \civ\ EQW and $\alpha_{\rm ox}$ can be seen much more clearly in the larger sample shown in Figure~4 of \citet{wvb+09}.  These trends between \civ\ blueshift, EQW, and $\alpha_{\rm ox}$ are expected given those already shown in Figures~\ref{fig:beff1550} and \ref{fig:c4b1550} and the known tight correlation between $L_{\rm UV}$ and $\alpha_{\rm ox}$ (which appears to be independent of redshift) \citep[e.g.,][]{ssb+06}.

\begin{figure}
\epsscale{1.0}
\plotone{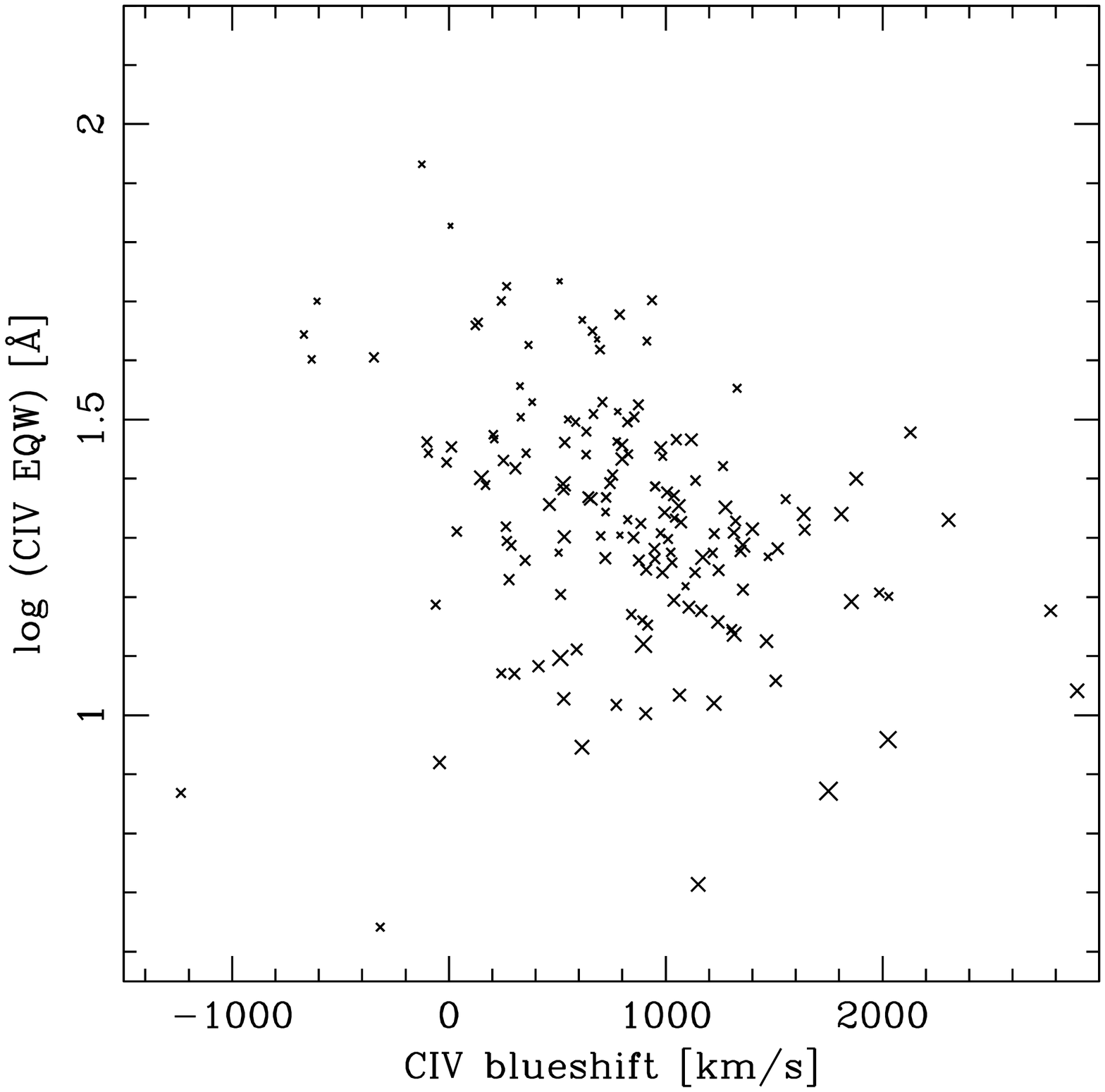}
\caption{\civ\ EQW vs. \civ\ blueshift.  Point size is a function of $\alpha_{\rm ox}$ with larger points indicating relatively weaker X-ray sources.  Note the relative lack of objects in the lower left-hand corner as compared with Fig.~\ref{fig:c4beqw}.  As with Fig.~\ref{fig:wu09}, $\alpha_{\rm ox}$ is taken from \citet{wvb+09}.}
\label{fig:wu09b}
\end{figure}

We can combine the \civ\ EQW, \civ\ blueshift, and $\alpha_{\rm ox}$ data into one plot as shown in Figure~\ref{fig:wu09b}.  
It is clear that the location of quasars in this 2-D parameter space is connected to the SED, with X-ray weaker objects occupying the lower right-hand corner.  The relative lack of objects in the lower left-hand corner may be indicative of intrinsic X-ray weakness in these objects, making them less likely to be detected in relatively shallow survey data. 

\subsection{Low-redshift Comparison}

What is perhaps most striking about Figures~\ref{fig:c4beqw} and \ref{fig:c4balphanu} is their similarity to the distribution of FWHM H$\beta$ vs.\ the strength of optical \ion{Fe}{2} from lower redshift quasars as shown in Figure~1 of \citet{szm+00} and to the distribution of  FWHM H$\beta$ vs.\ the soft X-ray spectral index as shown in Figure~8 in \citet{bbf96}.  In particular, all these plots show a missing quadrant rather than a correlation between the parameters.  In \citet{szm+00}, the RL quasars are also largely biased to the upper left-hand quadrant (weak \ion{Fe}{2} and broad H$\beta$).  This similarity should perhaps be no surprise as \citet{sbm+07} shows that their measure of \ion{C}{4} asymmetry (closely related to our blueshifts) is strongly correlated with \ion{Fe}{2} strength and that a plot of FWHM H$\beta$ versus \ion{C}{4} asymmetry shows a similar missing quadrant and RL population.  We will discuss the implications of this further in Section~\ref{sec:rlrq}.

Although \citet{rvr+02} also considered \ion{Mg}{2} and [\ion{O}{3}], we note that our high-redshift SDSS spectra do not have the spectral coverage to study \civ\ and H$\beta$ simultaneously in individual objects.  As such we have concentrated on the \civ\ line parameters in this paper; see \citet{Shen2010} for further discussion of the H$\beta$ and [\ion{O}{3}] lines.  However, we emphasize that the observed trends in \civ\ properties can be thought of as part of a more generic Eigenvector analysis \citep[e.g.,][]{bg92,szm+00}.  For example, the H$\beta$ line tends to be {\em redshifted} when \civ\ is close to systemic and that H$\beta$ has little shift (red or blue) when \civ\ is blueshifted \citep[e.g.,][]{msd+96,bl05}.  Weaker \civ\ would also seem to imply stronger \ion{Fe}{2} and weaker [\ion{O}{3}] in addition to a softer $\alpha_x$ and narrower H$\beta$ \citep[e.g.,][]{Wills99,Laor2000a}, though we emphasize that weaker \civ\ lines span the full range of observed blueshifts.

\section{Composite Spectra}
\label{sec:comp}


A wealth of diagnostic information comes from the various emission line features (particularly their ratios).  To increase the S/N of the emission features, one can stack individual spectra into a composite spectrum \citep[e.g.,][]{fhf+91,vrb+01}.  
However, as emphasized by \citet{bms+04} it can be particularly difficult to interpret composites made as a function of one parameter (e.g., $L$ in \citealt{dhs+02}, or even $L/L_{\rm Edd} \equiv L/M$ as was done by \citealt{whd04}).  Indeed, we are guilty of doing this as a function of \civ\ blueshift in \citet{rvr+02} and as a function of continuum color in \citet{rhv+03}; better would be to create composite as a function of both parameters.  

As such, we have constructed composite spectra broken in to small bins in parameter space to explore the overall spectral changes related to those parameters.  In particular, we parse the \civ\ EQW--blueshift plane (hereafter ``\civ\ parameter space'') into 9 bins as indicated in Figure~\ref{fig:c4beqw}.  We have chosen these two parameters as they are obviously both accessible in the same objects and have a large range of values.  Furthermore, their distribution is such that they apparently represent at least two underlying physical parameters; indeed, the \civ\ blueshift has been shown to be an EV1 correlate \citep{sbm+07}, while the \civ\ EQW may be related to EV2 (by way of luminosity and the BEff; \citealt[e.g.,][]{wlb+99}).   

\begin{figure}
\epsscale{1.0}
\plotone{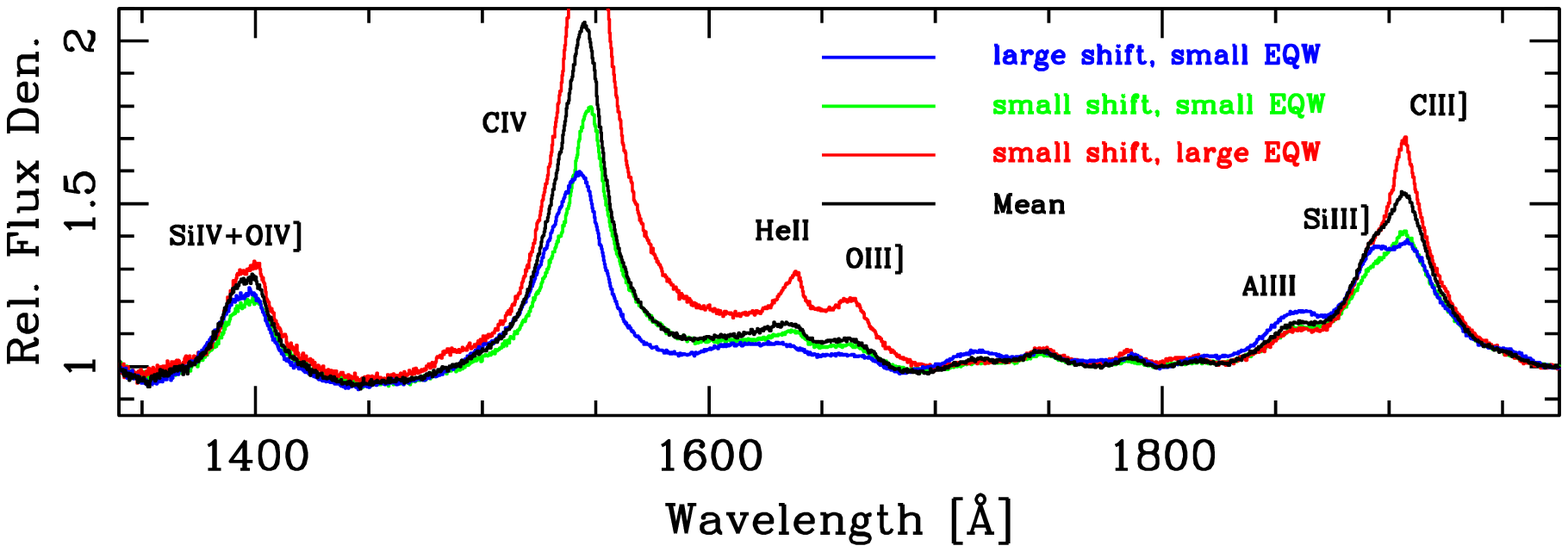}
\caption{Composite spectra from \ion{Si}{4}/\ion{O}{4}] through \ion{C}{3}] for 4 regions of \civ\ EQW-blueshift parameter space with $\sim$ 1000 spectra contributing to each composite; see Fig.~\ref{fig:c4beqw}.  Red indicates objects with large \civ\ EQW and small blueshift, blue indicates the opposite: small EQW and large blueshift.  Green indicates those quasars that fail to follow the expected trend with luminosity, namely small blueshift and small EQW, while black shows the ``mean'' spectrum consisting of objects that are intermediate in both parameters.  The spectra are normalized to unity at the ends of the plot region.}
\label{fig:lineplot3way}
\end{figure}

\begin{figure}
\epsscale{1.0}
\plotone{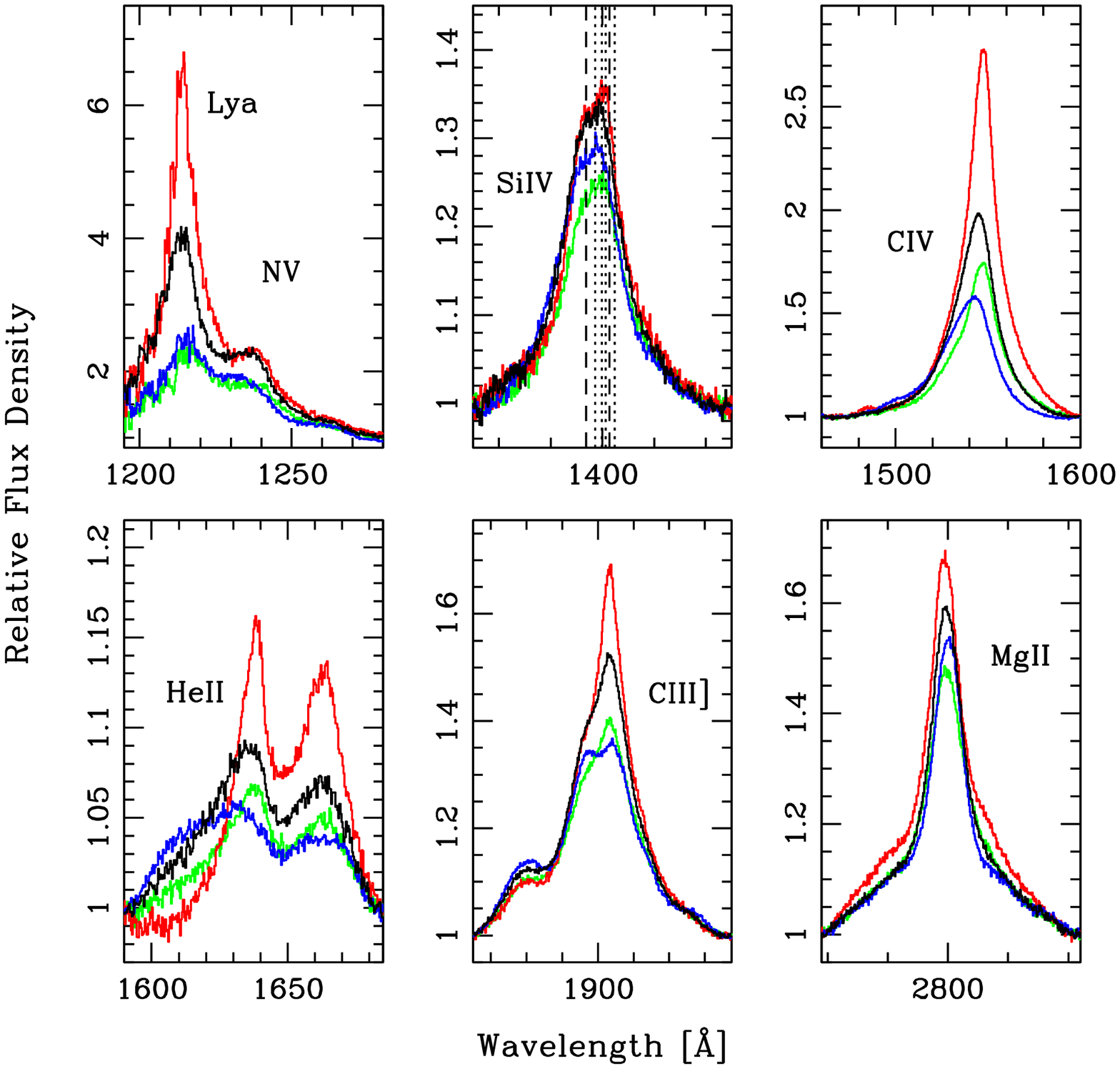}
\caption{As with Fig.~\ref{fig:lineplot3way} except zooming into the line features and also including Ly$\alpha$ and \ion{Mg}{2}.  The composites are normalized to unity at the edges of each panel (except the Ly$\alpha$ panel which uses the normalization of the \ion{Si}{4} panel as there are no continuum windows in the Ly$\alpha$ forest region). In the \ion{Si}{4} panel, the dashed vertical lines indicate the location of the doublet \ion{Si}{4} features, while the dotted lines indicate the five \ion{O}{4}] transitions.  In the Ly$\alpha$ panel many fewer quasars contribute to the composites than in the other panels.}
\label{fig:lineplot3waypanels}
\end{figure}

The divisions used to create the composites are shown by the dashed lines in Figure~\ref{fig:c4beqw} and were chosen to obtain roughly equal numbers of quasars ($\sim1100$) in the middle, upper-left, lower-left, and lower-right bins.  (Note that in \citet{rvr+02} only 200 quasars were included in each of the composites as a function of \civ\ blueshift only.)   We limit ourselves to exploring the composites in these 4 bins, using the other bins as a guideline to determine which trends are likely to be real.  In Figures~\ref{fig:lineplot3way} and \ref{fig:lineplot3waypanels} we show these four composites in two different presentations.  The first allows a comparison of weaker features between the lines and the relative continuum placement.   The second concentrates on the individual line regions, including the Ly$\alpha$+\ion{N}{5} blend and \ion{Mg}{2}.  It is important to note that these composites are are {\em not} created with luminosity-matched samples --- in part {\em because} the average luminosity changes by $\sim2$ dex across Figure~\ref{fig:c4beqw}.  However, as discussed in Section~\ref{sec:lme}, if the SEDs of the objects in the opposite corners of \civ\ parameter space are very different, it is difficult to create luminosity-matched samples.  If one matches on $L_{\rm UV}$ and the SEDs are different, then the samples will differ in $L_{\rm bol}$ and vice versa.  We do not, however, consider this a problem as we would argue that the shape of the SED is more important than any measure of the luminosity in determining the properties of the BELR.

\begin{figure}
\epsscale{1.0}
\plotone{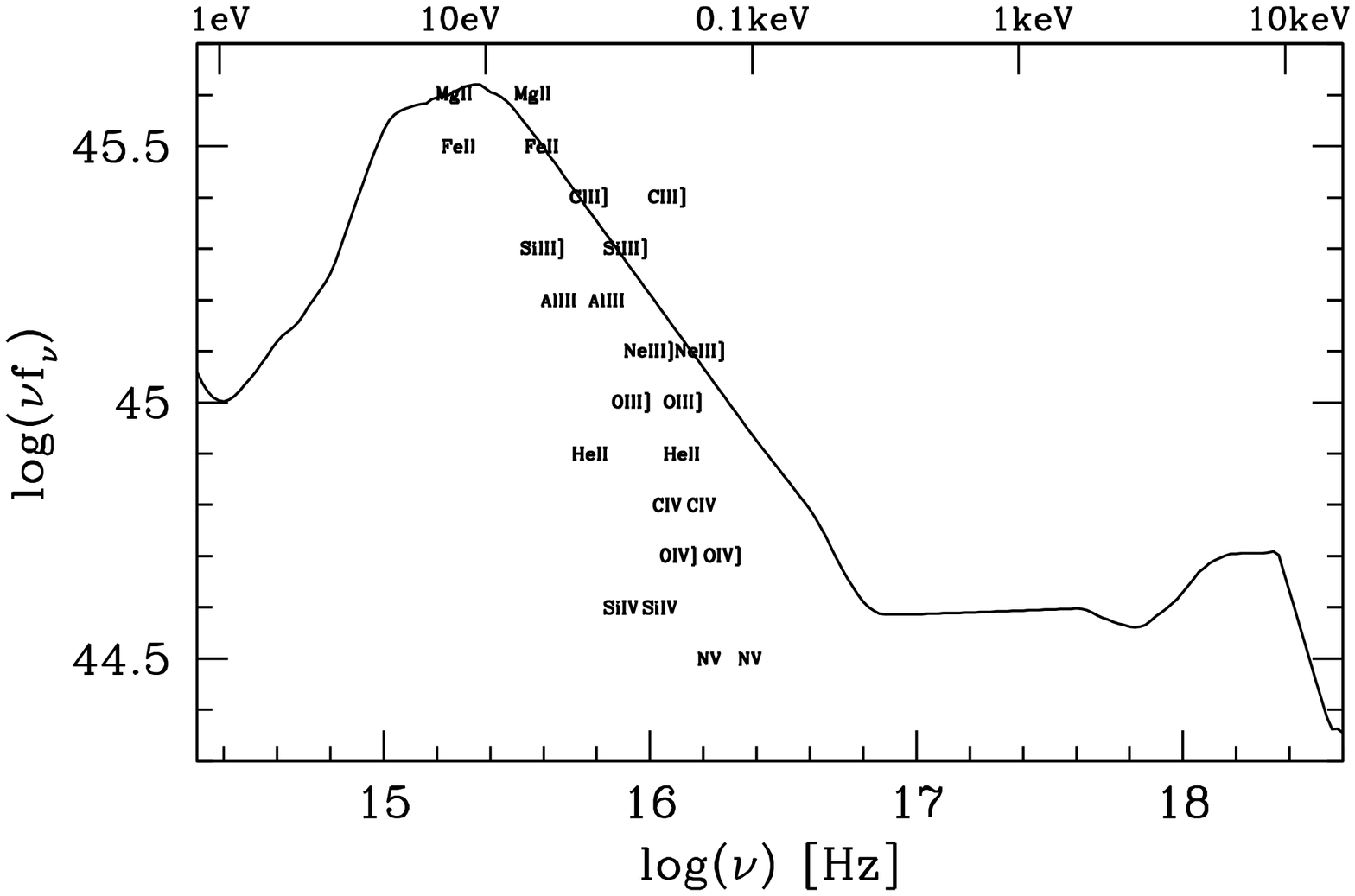}
\caption{Ionization potentials for the lines discussed herein.  Two values are given for each species.  The lower energy one is for creation, the higher energy for destruction.  Lines are organized vertically by wavelength --- there is no other meaning in their vertical placement.  We show the \citet{ewm+94} radio-quiet SED for comparison.  Ly$\alpha$ emission is at $10.2\,{\rm eV}$.}
\label{fig:ipplot}
\end{figure}

The most obvious differences between the composite spectra are seen in the \civ\ line --- by construct, of course.  In addition to the trends expected from our choice of binning, a few other trends are evident.  We consider these from long to short wavelength.  In considering these lines, it is helpful to know their ionization potentials (for both creation and destruction); these are illustrated in Figure~\ref{fig:ipplot}.

We start with the \ion{C}{3}] $1909\,{\rm \AA}$ complex, where \ion{C}{3}] is blended with \ion{Si}{3}] $1892\,{\rm \AA}$ and \ion{Al}{3} $1857\,{\rm \AA}$.  As noted by \citet{wlb+99} the \ion{Si}{3}]/\ion{C}{3}] ratio is generally considered a good indicator of density and is one of the primary Eigenvector 1 correlates. This is especially interesting given the argument put forward by Morris et al.\ (1993) and van Gent (2001) that the Baldwin effect seen in Wolf-Rayet stars can be attributed to density effects \citep{Shields07}.  For example, van Gent et al. (2001) argue that the BEff is seen in W-R stars, but not normal O stars because the atmospheres of normal O stars are not dense enough.  However, \citet{clb06} have shown that the \ion{Si}{3}] 1892/\ion{C}{3}] 1909 ratio can instead be an indicator of the SED.  They note that \ion{Al}{3} can be used to determine whether this ratio is a density or SED indicator.  In the case of our large \civ\ blueshift composite, which has the largest \ion{Si}{3}]/\ion{C}{3}] ratio, we find that \ion{Al}{3} is also particularly strong.  As such, the SED dependence of this line ratio must be significant and a high \ion{Si}{3}] 1892/\ion{C}{3}] 1909 ratio is indicative of an X-ray weak spectrum for large \civ\ blueshift quasars.  This is perhaps not surprising given the trends seen in Figure~\ref{fig:wu09b}.
However, it is only for our large blueshift composite that \ion{Si}{3}] is nearly as strong as \ion{C}{3}] and \ion{Al}{3} is so strong (and not for weak, unblueshifted objects).  Thus it seems that the \ion{C}{4} blueshift is strongly associated with a relatively weak X-ray spectrum --- independent of the strength of the emission lines (and the BEff, in general).  An SED relatively weak at X-rays suggests strong line driving by radiation pressure as it means that $L_{\rm UV}$, which drives the wind, is high relative to $L_X$, which can suppress a wind through over-ionization.  As such, the \ion{C}{3}] complex seems to support the general idea of the BELR being the combination of disk and wind components, with one dominating over the other depending on the shape of the SED \citep[e.g.,][]{Leighly04}.

In reality a wind may always be present, but may be driven by
something other than radiation line pressure in the ``disk'' systems.
In addition, the wind's parameters (launching radius, opening angle,
terminal velocity, etc.) may change as a function of the SED.  As
such, when we talk about a trade-off between the disk and wind components,
what we mean specifically is that the ``wind'' objects have
optimal conditions for a radiation line driven wind and that the
``disk'' objects have a wind that allows more of the ionizing
continuum to reach the part of the BELR where the disk component
is formed.

Moving blueward towards \civ\ there is the blend of \ion{He}{2} $1640\,{\rm \AA}$ and \ion{O}{3}] $1663\,{\rm \AA}$.  Both of these lines are extremely strong in the large \civ\ EQW composite.  The \ion{He}{2} line is strongly asymmetric and blueshifted in both the mean and large \civ\ blueshift composites, whereas the \ion{O}{3}] appears similarly reduced in strength, but not obviously blueshifted or asymmetric.  The behavior for \ion{He}{2} is not surprising in that strong \ion{He}{2} is suggestive of a stronger (harder) soft X-ray spectrum \citep{Leighly04} and given our finding above for the \ion{C}{3}] complex.  A strong soft X-ray spectrum would inhibit a wind; whereas a weaker soft X-ray continuum would allow a strong wind to develop \citep{Leighly04} --- again supporting the notion of a disk/wind trade-off with SED.  

For \civ\ itself, other than the properties that are seen given the
subsample definitions, we note that for the low-blueshift, low-EQW
objects, there is an asymmetry in the blue wing of \civ\ which may be
indicating that there is a significant wind component in these systems as well,
but with {\em both} the disk and wind components being weak (and of
relatively equal strength).  On the blueward side of \civ, there appears to be an excess of \ion{N}{4}] $1486\,{\rm \AA}$
  emission in the large \civ\ EQW composite.  This feature is
  relatively rare in quasars and is thought to be indicative of
  enhanced metallicity \citep{bho04}.  This finding is interesting with respect to the results of \cite{Leighly04} where CLOUDY modeling of two narrow line Seyfert 1 galaxies with unusual SEDs yielded a best fit using a model with enhanced metallicity.

Moving blueward, as with many other authors, we find that the $1400\,{\rm \AA}$ feature does not seem to participate in either the BEff or \civ\ blueshift trends.  \citet{Leighly04} have interpreted this as being due to a trade-off between one line (\ion{Si}{4}) having a strong disk component and the other (\ion{O}{4}]) having a strong wind component.  Indeed, this interpretation is consistent with the SED results that we have derived from the other lines.  Moreover, as our large \civ\ EQW composite shows evidence for a strong \ion{Si}{4} contribution in the form of a resolved doublet feature in the emission line, it is clear that \citet{Leighly04} have properly interpreted \ion{Si}{4} as being associated with the disk component.  We emphasize that our use of redshifts as corrected by HW10 are what allow this level of spectral resolution in the composite spectra.  

In general, the Ly$\alpha$ region in Figure~\ref{fig:lineplot3waypanels} should be viewed with caution as the $z\le2.2$ limit on the sample means that far fewer quasars contribute to this region of the composite spectra.  However, the ``middle bin''  composites not shown in Figure~\ref{fig:lineplot3waypanels} would seem to support the trend of 
increasing Ly$\alpha$ EQW with increasing \civ\ EQW.  
We note that the inflection between Ly$\alpha$ and \ion{N}{5} is less pronounced in those objects with large \civ\ blueshift.  The lack of an inflection in the strong blueshift composite strongly suggests that the \ion{N}{5} line is also blueshifted in these objects and has a strong wind component.  This shift is interesting in that the \ion{N}{5} line does not follow the BEff trend \citep[e.g.,][]{dhs+02}.  Investigations of metallicity using this line require great caution when deblending \ion{N}{5} from Ly$\alpha$ (as do investigations of the BEff in \ion{N}{5}).

In the disk+wind scenario these composite spectra paint a compelling picture in terms of the SED dependence of the relative contributions of the disk and wind.  In the composite with a strong ``disk'' component a strong ionizing spectrum is needed to produce the strong UV emission lines that have ionization potentials near $\sim50\,{\rm eV}$.  Such a spectrum will also inhibit a radiation line driven wind as these hard photons will ionize the atoms that would otherwise constitute the wind --- disrupting the radiation line driving force.  For objects that are relatively weaker in the X-ray, a strong radiation line driven wind can form and said wind will further reduce the hard X-ray flux that reaches the disk component as the wind will ``filter'' the continuum \citep{Leighly04,lc07}.  Such objects will also be more likely to be seen as being absorbed (and not just intrinsically weak) at X-ray energies.  This naturally accounts for the parent population of BAL quasars; see Section~\ref{sec:balqsos}.

\section{Discussion}
\label{sec:discussion}

\subsection{Radio-Quiet vs.\ Radio-Loud}
\label{sec:rlrq}

These data can be used to address the issue of the radio-loud/radio-quiet dichotomy, particularly within the EV1 context.  This discussion follows from a long history, extending from \citet{bg92} and \citet{bor02} to \citet{sbm+07} and beyond.

\begin{figure}[h]
\epsscale{1.0}
\plotone{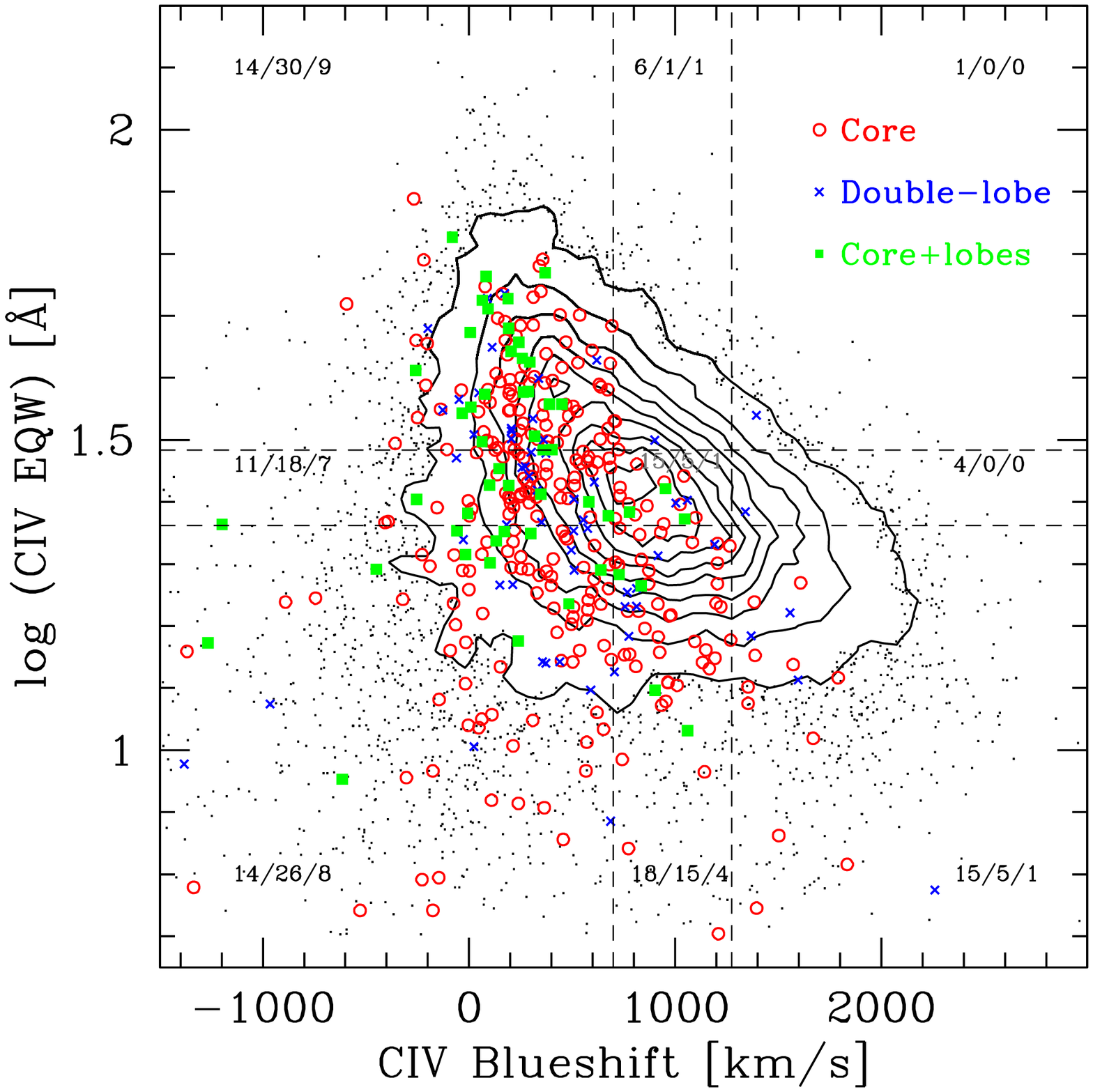}
\caption{As in Fig.~\ref{fig:c4beqw}.  RL objects here are additionally divided by morphology with core sources (flat-spectrum) in red, double-lobe sources in blue, and core+lobe sources in green.  In each of the 9 sections, three numbers indicate:  1) the percentage of all RQ quasars that are located in that bin, 2) the percentage of all RL quasars that are located in that bin, and 3) the percentage of RL quasars in that bin (as compared to all quasars in that bin).  For example the high-blueshift, low-EQW quasars in the lower right-hand corner account for 15\% of RQ quasars in the sample and 5\% of the RL quasars in the sample, but only 1\% of the quasars in that corner are RL (99\% are RQ).  The values in the central bin are $15/5/1$.  Restricting the sample to high S/N objects does not change the distribution of radio morphologies.}
\label{fig:c4beqwmorph}
\end{figure}

It is particularly interesting to consider the RL dichotomy in the context of Figure~\ref{fig:c4beqw}, which we have reproduced in Figure~\ref{fig:c4beqwmorph} with additional information regarding the RL properties.  Other than the radio-loudness, \civ\ EQW and blueshift are the two quasar parameters that show the largest dynamic range for high-redshift quasars.  Moreover, the \civ\ blueshift has been shown to be correlated with the relative strength of (optical) \ion{Fe}{2} \citep{sbm+07}, where \ion{Fe}{2}/[\ion{O}{3}] and FWHM H$\beta$ are the two low-redshift parameters with the largest dynamic range and are dominant in PCA decompositions.  As such, if RL quasars were distinct from RQ quasars, it would be reasonable to expect that they would be distinct in the \civ\ parameter space and if orientation played a significant role in the observed BELR parameters, then we might expect lobe-dominated (steep-spectrum) RL and core-dominated (flat-spectrum) RL quasars to separate in \civ\ parameter space.

Indeed, \citet{bor02} found that RL quasars occupy an extrema of EV1-EV2 parameter space --- even after removing radio properties from the PCA analysis.  In earlier work \citet{bg92} found that there were few RQ counterparts to steep-spectrum RL quasars and argued that the RL objects represented an extrema in the quasar population (rather than a parallel sequence) with orientation (as determined from the radio spectral index) not being the dominant contributor to the location of quasars in EV1-EV2 parameter space (see also \citealt{smd00}).

In terms of \civ\ parameter space, our results would seem to agree with the notion that orientation does not play a significant role.  We find that RL objects with core, double-lobe, and core+double-lobe morphologies all span a similar space in terms of their \civ\ EQWs and blueshifts; see Figure~\ref{fig:c4beqwmorph}.  Assuming that these morphologies represent different orientations (in the ensemble average), it must be that orientation is not the major factor in determining the observed properties of the \civ\ BELR.   Kimball et al.\ (2011, submitted) reached a similar conclusion based on an independent analysis.  In reality, the differences between Type 1 and Type 2 AGNs tells us that orientation {\em does} matter; it simply appears to be of lesser importance than other parameters (at least over the range of opening angles for Type 1 quasars).

While the radio morphologies do not appear to show the expected dependence on orientation, we do find significant differences between RL and RQ quasars in the \civ\ parameter space.  In particular, while the RL quasars do not fully span the parameter space occupied by the RQ quasars, they also do not occupy a distinct part of the parameter space (contrary to the results of \citealt{bg92} and \citealt{bor02}).  However, RL objects are strongly biased towards small \civ\ blueshifts (and, to a lesser extent, large \civ\ EQWs) and high blueshift quasars are almost exclusively RQ.  This can be seen clearly by examining the RL and RQ percentages in each part of the parameter space shown in Figure~\ref{fig:c4beqwmorph}.  Fully 30\% of RL quasars occupy the region with the largest \civ\ EQW and smallest blueshifts, whereas only 14\% of RQ quasars occupy that same part of parameter space. Yet the RQ quasars are still dominant --- making up 91\% of the quasars in that corner of the distribution (9\% being RL).    Sulentic and collaborators find a similar distribution of RL and RQ quasars in \ion{Fe}{2}/[\ion{O}{3}] vs.\ FWHM H$\beta$ parameter space and have argued for a two population model that subsumes the RQ/RL distinction \citep[e.g.,][]{sbm+07}.

\citet{szm+00} suggest that the RL-RQ split argues for a division of quasars into two populations: one dominated by RL quasars with some RQ quasars and one dominated by RQ quasars.  Our data do not support such a binary distinction, but the general point is valid nevertheless.  That is, there does seem to be a parameter space that is uniquely occupied by radio-quiet quasars at one end, with radio-loud quasars occupying the other end of the (apparently continuous) distribution, and some radio-quiet quasars also inhabiting the RL parameter space.  

We see the data as evidence for continuous changes in a 2-D (or more) parameter space where the SEDs (and thus wind structure and line parameters) are continuously morphing between extrema.  For example, radiation line driving may dominate in the ``wind'' objects, while the ``disk'' objects perhaps have MHD-driven winds \citep{Proga03,Richards2006}.  In the context of our adopted disk+wind model, we interpret this as meaning that the RL quasars have a stronger disk component than RQ quasars as a whole, but that some RQ quasars have more RL-like disk components.  For both the RL and RQ quasars, the radiation line driven wind component of the BELR appears to increase with increasing UV luminosity; it may be that the RL quasars with large \civ\ blueshift simply have softer SEDs than the majority of RL quasars and have a larger radiation line driving contribution to their winds.

\begin{figure}
\epsscale{1.0}
\plotone{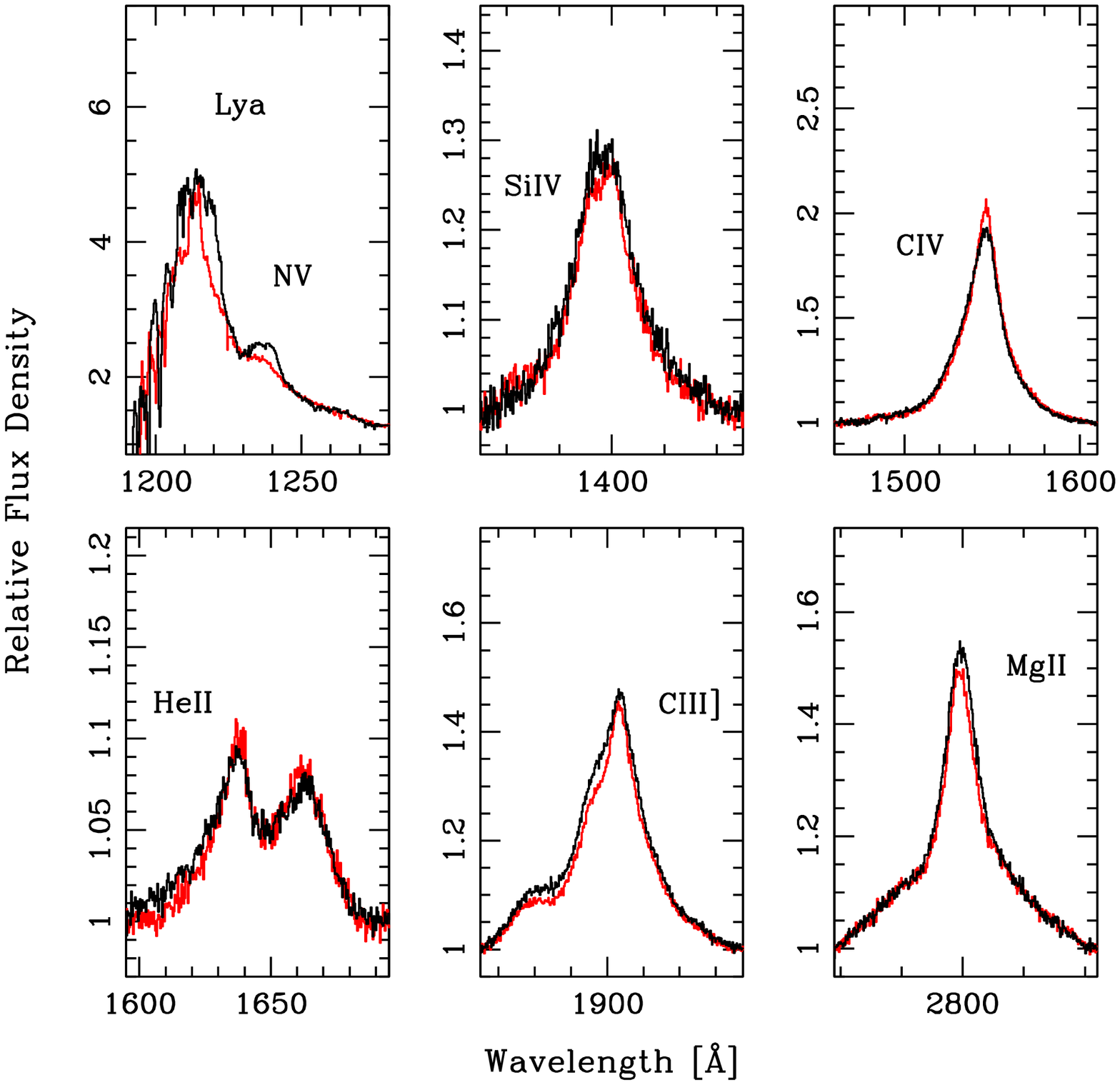}
\caption{Composites of RL quasars (red) and RL-like RQ quasars (black).  Objects are matched on luminosity, redshift and \civ\ emission line parameters.  The good agreement between the other lines (as compared to the range seen in Figure~\ref{fig:lineplot3waypanels}) suggests that the BELR (and thus the SED) is similar in the RL and RL-like RQ quasars. }
\label{fig:lineplotrlrq}
\end{figure}

Is it possible then that the only distinction between the RL quasars and those RQ quasars that occupy the same parameter space as RL quasars is their radio emission?  To test this notion, we have created matched samples of RL quasars and ``RL-like'' RQ quasars; the resulting composite spectra are shown in Figure~\ref{fig:lineplotrlrq}.  The samples were matched on \civ\ EQW, \civ\ blueshift, luminosity, and redshift to within $0.8\sigma$ for each of those parameters; the matching range was chosen empirically and minimizes the number of RL objects for which there is no match (4 of 369) while still providing a tight matching tolerance on each of these five parameters.  Although only the \civ\ line parameters were used in the matching,  most of the the other emission line features are essentially identical to within the errors (the differences are much smaller than the ranges spanned by the full sample; compare to Fig.~\ref{fig:lineplot3waypanels}).  Both the RL and RQ objects were restricted to $i<18.9$, where a non-detection in FIRST makes an object formally RQ \citep{imk+02}.  Even if we restrict the RL sample to the top 10\% of the radio luminosity distribution, we can still find RQ quasars that provide suitable matches (at least in the ensemble average).  Thus, there is little distinction in the BELR properties of the RL quasars and RL-like RQ quasars; they are distinguished only by their radio properties.  In terms of the BELR, it seems that what distinguishes a RL (and RL-like) quasar is its strong X-ray flux (relative to UV) and the resulting strong disk component.  

We note that the apparent similarly of some RQ quasars to RL quasars is the reason that we have not opted to use the redshift corrections specific to RL quasars as tabulated in HW10.  These corrections are based on the fact that RL quasars have different emission lines shapes than RQ quasars on average (as we have seen here).  Yet, at least some RQ quasars have very similar emission line shapes as RL quasars.  As such, it would not be appropriate to adjust the RL redshifts without similarly correcting the redshifts of the RL-like RQ quasars.  Had we done so, there would have been a small shift in the peaks of the lines between the RL and RL-like RQ composites in Figure~\ref{fig:lineplotrlrq}.

As for why \citet{bor02} find that RL quasars are an extrema distinct from RQ quasars, it is worth noting that the ``PG'' quasar sample used in \cite{bg92} is known to be biased towards RL quasars; the answer to our differing results may be hidden in the answer to that bias.  In addition, \citet{bor02} supplemented the PG sample with 75 RL quasars that may not be completely comparable to the PG sample.  Lastly, our results don't disagree that RL quasars are extrema in some sense -- rather our results suggest that some RQ quasars do indeed have similar properties (see also \citealt{szm+00}).

Some authors have argued that RL quasars appear to be more massive, on average, than RQ quasars \citep{Laor2000b,llr+01,mj04} and \citet{bor02} has argued that high mass is what separates the RL and RQ quasars in EV1-EV2 parameter space.  If that is the case, then our finding that some RQ quasars occupy the RL part of \civ\ BELR parameter space may be an argument that mass may be a necessary, but insufficient condition for RL.  The most likely secondary parameter is black hole (BH) spin \citep{bla90,wc95,Sikora07}; in the ``spin paradigm'' high black hole spin is needed to form radio jets.  RL-like RQ objects could be explained as high-mass quasars that have insufficient spin, perhaps as a result of mass accretion onto an (initially) retrograde spinning BH. Overall, RL quasars would be expected to be more massive (and higher spin) if black holes on average get ``spun up'' as they grow in mass \citep[e.g.,][]{wc95,up95}.  

In terms of morphologies, since elliptical galaxies host both RL and RQ quasars, while spiral galaxies host only RQ quasars
\citep[e.g.,][]{dmk+03}, we might expect that the RL-like RQ objects will have different morphology than the remainder of the RQ sample.  Specifically, a testable prediction is that those RQ quasars that are hosted by massive elliptical galaxies will have RL-like BELR properties (with smaller BH spins than their RL counterparts). Another test that we plan to perform is to see if the ensemble average (from stacking analysis) radio flux of RL-like RQ quasars and those RQ quasars without RL counterparts in \civ\ parameter space are different.

\subsection{BALQSOs}
\label{sec:balqsos}

\begin{figure}
\epsscale{1.0}
\plotone{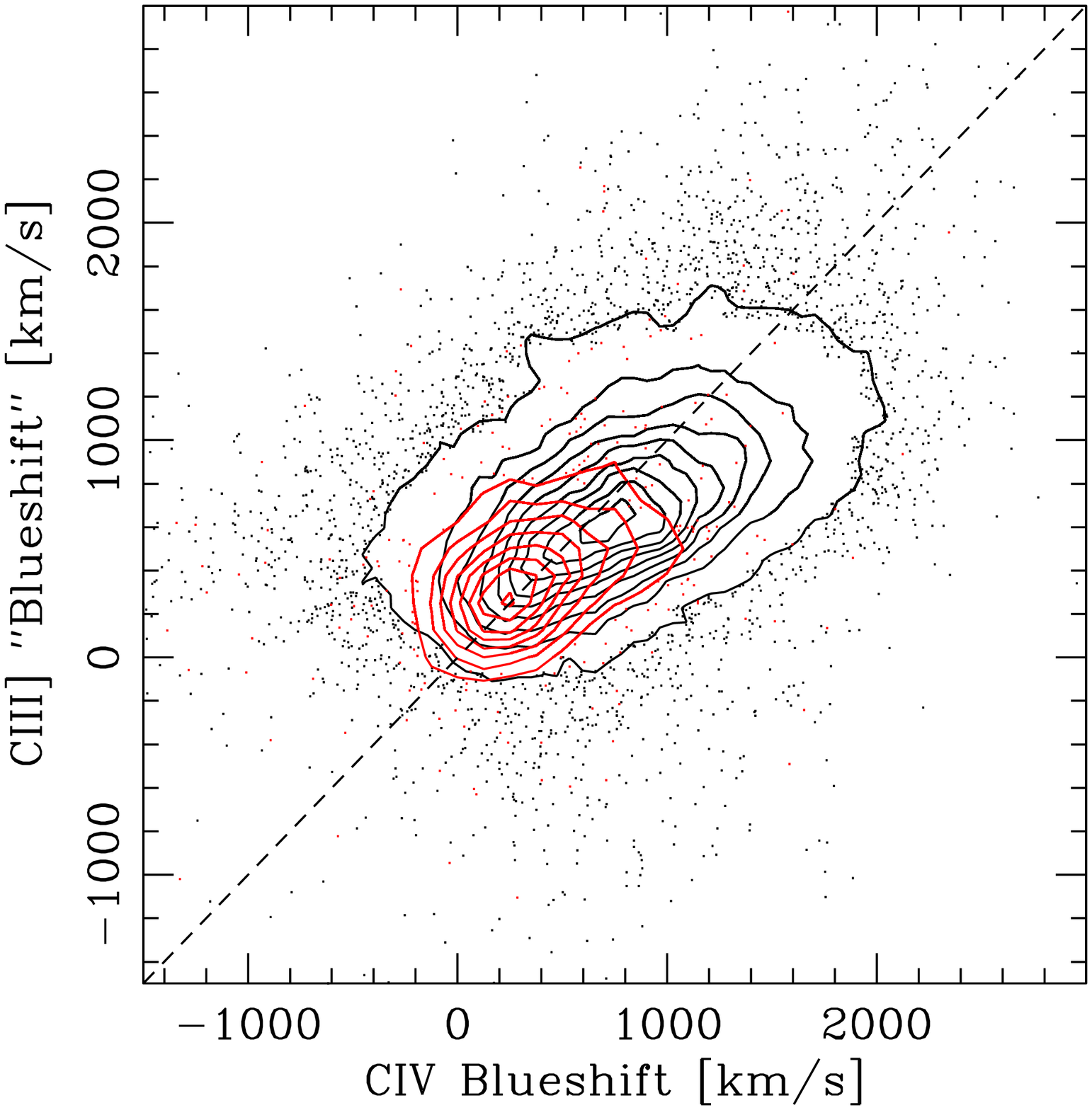}
\caption{\civ\ blueshift vs.\ \ion{C}{3}] ``blueshift''.  RQ quasars are shown in black and RL quasars in red.  The blueshift of the \ion{C}{3}] line is not an actual shift of the line centroid as it is for \civ\, but rather reflects the relative strength of \ion{C}{3}] vs. \ion{Si}{3}] since the SDSS pipeline does not deblend the two lines.   The strong correlation between these two blueshifts in RQ quasars suggests that the measurement from \ion{C}{3}] can be used as a rough surrogate for \civ.   The dashed line is a line of equality; it is not a fit to the data.}
\label{fig:c4bc3b}
\end{figure}

It is difficult to investigate broad absorption line quasars (BALQSOs; \citealt{wmf+91}) in the \civ\ emission line parameter space as the line emission is often partially absorbed by the absorption line gas.  However, it is possible to use the \ion{C}{3}] line complex as a surrogate.  Specifically, for the SDSS quasars, the spectroscopic pipeline does not deblend the \ion{Si}{3}] from the \ion{C}{3}] line.  Thus the measurements in the database are essentially a composite of the two lines.  As it happens, this hybrid line has a ``blueshift'' that tracks the \civ\ blueshift quite well as can be seen in Figure~\ref{fig:c4bc3b}.  Here the \ion{C}{3}] ``blueshift'' is actually due to the relative flux of the \ion{C}{3}] and \ion{Si}{3}] emission lines which the SDSS spectroscopic pipeline does not deblend.  Quasars with stronger \ion{Si}{3}] have larger \ion{C}{3}] ``blueshifts''.

\begin{figure}
\epsscale{1.0}
\plotone{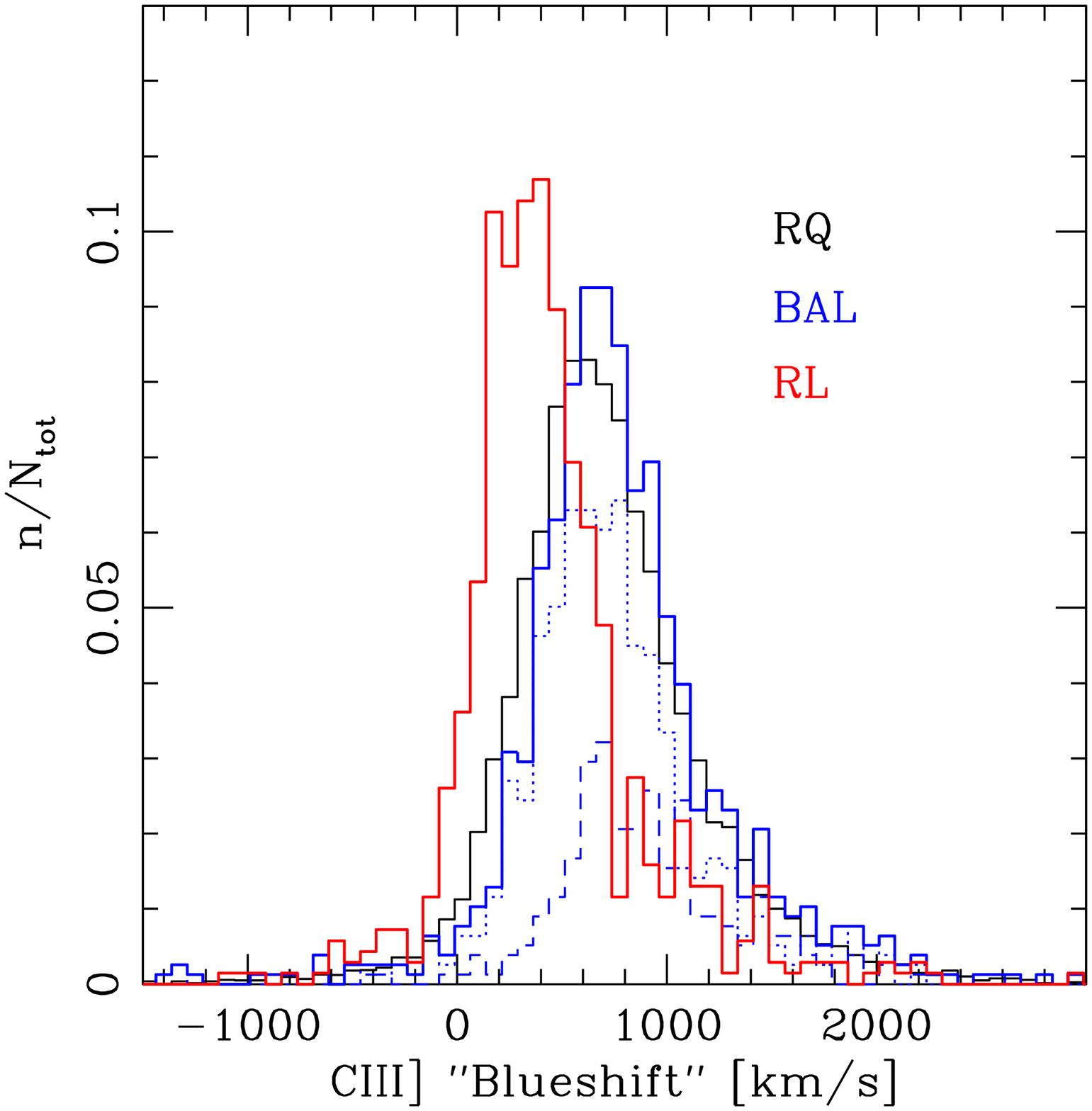}
\caption{Distribution of \ion{C}{3}] ``blueshifts'' for RQ ({\em black}) quasars, RQ BAL quasars ({\em blue}), and RL quasars ({\em red}).  The $y$-axis is scaled by the total number of objects in each category.  The BALQSOs are further split into objects with balnicity index, BI, greater than $2000\,{\rm km\, s^{-1}}$ ({\em dashed blue line}) and BI$\le2000\,{\rm km\, s^{-1}}$ ({\em dotted blue line}) as given by \citet{Allen10}; these two histograms are normalized by the total number of BALQSOs (i.e., the dashed and dotted blue lines sum to the solid blue line).}
\label{fig:c3bhist}
\end{figure}

In Figure~\ref{fig:c3bhist} we see that BALQSOs have somewhat larger \ion{C}{3}] blueshifts (and, by inference, larger \civ\ blueshifts) than RQ quasars in general.  Furthermore, the strongest BALQSOs, having ``balnicity indices'' \citep{wmf+91} larger than $2000\,{\rm km\, s^{-1}}$ have even larger average blueshifts.  While the scatter in Figure~\ref{fig:c4bc3b} is large, we note that the \ion{C}{3}] blueshift under-predicts the \civ\ blueshift for large \civ\ blueshift objects.  As such, the separation between RL quasars and BALQSOs seen in Figure~\ref{fig:c3bhist} actually must be under-estimated.
Thus it is clear that RL and BAL quasars are not drawn from the same distribution, with very strong BALQSOs being particularly unlikely to be RL \citep[e.g.,][]{smw+92,hf03,sds08}.  In the accretion disk wind model of \citet{mcgv95} the radio-loud quasars lack strong BAL troughs because of their strong X-ray flux, which inhibits the formation of a strong radiation line-driven disk-wind.

These findings are quite consistent with our results from \citet{rrh+03} where it was found that BALQSOs have larger than average blueshifts and smaller than average \civ\ EQWs (and bluer colors after correcting for dust reddening), with LoBALs having the most extreme properties --- by virtue of analysis of BALQSO composite spectra outside the \civ\ region.  This is also consistent with our finding above that large \civ\ blueshift objects have SEDs that allow the wind component to dominate over the disk component.

The fact that we can identify a parent population of BALQSOs among those (Type 1) RQ quasars without BAL troughs (those quasars with large \civ\ blueshifts due to a strong radiation line driven wind), coupled with the result of normal IR properties of BALQSOs \citep{ghb+07} means that BALQSOs cannot, in general, be a small population of completely dust enshrouded quasars as some authors have suggested \citep[e.g.,][]{bm92,cs01,ubw+09}.  This is important in the context of the observed BAL fraction of $\sim20$\% \citep[e.g.,][]{hf03}.  We argue that only quasars with strong radiation line-driven winds can have BAL troughs (seen only along certain lines of sight).  Those quasars with strong X-ray spectra and weak wind components might even have a nearly 0\% BAL covering fraction (though narrow absorption is not excluded), while those quasars with weaker X-ray spectra spectra and a strong wind component would then have a much larger than 20\% BAL covering fraction to produce an average of 20\% over the entire quasar population.  Thus the notion of BALQSOs being due to either orientation {\em or} evolution seems misguided at best.  Objects with different wind structures will have larger/smaller BAL covering fractions, which means that orientation is important.  At the same time, the structure of the wind appears to be SED dependent, which, in turn must be dependent on evolutionary parameters such as mass and accretion rate.  Thus in a disk-wind scenario driven by the SED, it is difficult to avoid having both orientation and evolution contributing to the observed BAL properties of quasars \citep{Richards2006,Allen10}.

\subsection{Reflections on Richards et al.\  (2002)}

As it has been eight years since our initial work on the matters discussed herein using the first SDSS data available, it seems appropriate to comment on how the field has evolved and how our thoughts have changed on some things since the publication of \citet{rvr+02}.

First, in \citet{rvr+02} we suggested that the \civ\ blueshifts were somehow
related to quasar orientation, whether external orientation (i.e., our
line of sight), or internal orientation (e.g., the opening angle of
the disk-wind).  Here we would agree with \citet{Leighly04} that the
differences in \ion{He}{2} emission-line properties of quasars with
large and small blueshifts are not indicating differences in external
orientation, but rather differences in the SED.  We do, however,
expect that these different SEDs will translate to differences in the
wind structure (and possibly the opening angle of the disk-wind).  

Second, we suggested that the \civ\ blueshift might result from a preferential reduction/suppression/obscuration of the red wing of the \civ\ emission line.  This suggestion has often been interpreted as meaning that the redshifted emission-line gas is obscured from our line of sight.  While this could be the case, in the model that we consider herein, the blueshift results from the outflowing wind component being stronger than the disk component.  That said, our real fear in \citet{rvr+02} was that, since the disk-wind model of \citet{mcgv95} produces a single-peaked C IV emission line through radiative transfer effects, such effects might also alter the symmetry of the line profile.  Even within the disk+wind picture this is still a concern; Hall et al.\ (in preparation) are considering this matter further.

\subsection{Luminosity, Mass, and Eddington Ratio}
\label{sec:lme}

The tendency is to assume that all quasars are deviations from a single mean SED (e.g., \citealt{ewm+94}) --- despite warnings from those authors that the range is large.  We then generally assume that it is possible to scale from a monochromatic luminosity to the full bolometric luminosity using a single scaling factor (i.e., the ``bolometric correction'').  While our own work shows that the bolometric corrections can differ between objects by only a factor of $\sim$2 \citep[e.g.][see also \citealt{gkl+10}]{rls+06}, it is important to consider whether the range of bolometric corrections is systematic in any way (e.g., across the \civ\ parameter space).  Our findings would seem to argue that the SED in the ionizing part of the SED must be very different as one moves from the upper left-hand corner of Figure~\ref{fig:c4beqw} to the lower right-hand corner.  Based on the BELR trends seen in Section~\ref{sec:c4prop}, those quasars with strong disk components are likely to have more ionizing flux and thus larger bolometric corrections than are generally assumed.  Moreover, it is important to realize that it is generally not possible to make measurements of quasar SEDs in a crucial portion of the ionizing continuum due to the far-UV opacity of the Universe, which leads to further uncertainty in the bolometric correction.  

Even if bolometric luminosities are currently being estimated accurately, the issue of spin has important implications for estimating the accretion rate from the bolometric luminosity.  In particular, $L_{\rm
  bol}$ yields an estimate of $\dot{M}$ given that $L_{\rm
  bol} = \eta\dot{M}c^2$.  However this assumes that all quasars have black holes that have the same radiative efficiency, $\eta$.  In reality, it is quite possible that quasars that occupy different extrema in the \civ\ emission line parameter space have very different $\eta$ values.
Specifically, if RL quasars have higher spin than RQ quasars (on average), then they will have higher efficiencies, which translates to smaller accretion rates (at a fixed bolometric luminosity).   If that were the case, then two objects in opposite corners of Figure~\ref{fig:c4beqw} with the same bolometric luminosity could have very different accretion rates.

\begin{figure}
\epsscale{1.0}
\plotone{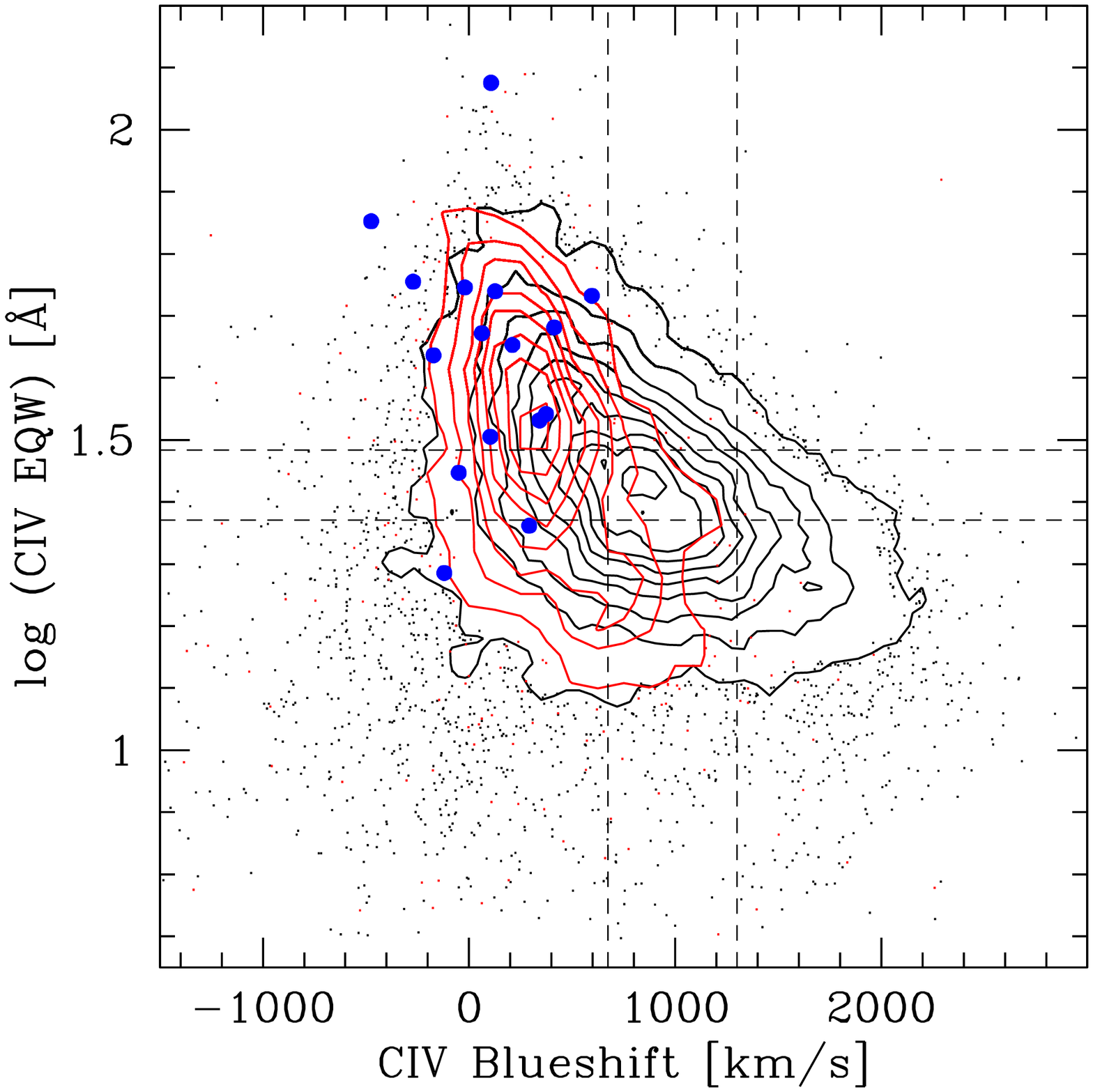}
\caption{As in Fig.\ref{fig:c4beqw}, including the distribution of \civ\ EQW and blueshifts for the reverberation mapped quasars from \citet{ksn+00} (filled blue circles).  \civ\ EQW values for the \citet{ksn+00} objects are from \citet{bl04}; \civ\ blueshift values are derived from \citet{bl05} and \citet{sbm+07}.  These objects do not span the full parameter space, calling into question the accuracy of the BH mass estimates for objects that occupy other parts of the \civ\ emission line parameter space.  The 0.2 dex under-estimate of \civ\ EQW by the SDSS pipeline has no affect on lack of large \civ\ blueshift objects in the reverberation mapping sample.}
\label{fig:kaspi}
\end{figure}

In terms of black hole mass scaling relations, we emphasize that current scaling relations \citep[e.g,][]{vp06} are based on a small number of reverberation mapped objects including just 17 objects from \citet{ksn+00}.  That being the case, we point out a previously unreported observation: the \citet{ksn+00} quasars occupy only part of the \civ\ emission line parameter space; see Figure~\ref{fig:kaspi}.  As such, the scaling relations derived from these results may not be applicable to quasars outside of that part of parameter space (certainly for \civ-based masses).  A significant observing campaign targeting quasars that span the full \civ\ parameter space will be needed to test the validity of extrapolating current scaling relations into other parts of \civ\ parameter space and to investigate any possible systematics in H$\beta$- and \ion{Mg}{2}-based mass estimates (e.g., two objects with the same $L_{\lambda}$ that have very different ionizing SEDs will have very different wind properties and could have different H$\beta$ and/or \ion{Mg}{2} BLR sizes, leading to a systematic error in determining $R_{BLR}$ from $L_\lambda$ alone).  Such an effort may prove to be difficult if current studies are biased towards objects for which it is easiest to perform reverberation mapping.  In the interim, it is clear that we must be aware of possible systematic errors in mass estimates for quasars with strong disk winds.

Errors in $L_{\rm bol}$ and $M$ will propagate to the Eddington ratio, $L_{\rm bol}/L_{\rm Edd}$, which is generally computed from $L_{\rm UV}/M_{\rm scaling}$.  
However, we caution against focusing on the Eddington ratio alone,
arguing that the simultaneous specification of the black hole mass and
the accretion provides greater insight. 
The fundamental problem is this: while it is true that objects with high accretion rate and low mass will have large Eddington ratios, and that objects with low accretion rate and high mass will have small Eddington ratios, it is {\em not
  necessarily true} that objects with intermediate Eddington ratios can be treated the same regardless of their masses.  Both high accretion rate and high mass, along with low accretion rate and low mass can yield the same (intermediate) Eddington ratio.  However, those objects can have very different properties.  For example if there is a mass-dependence to the creation of RL quasars, then it could be dangerous to treat RL and RQ quasars with the same Eddington ratio distribution as being comparable without also knowing that their mass distributions match.

In short, understanding \civ\ parameter space has important implications for estimating the accretion rates and masses of quasars and further work is needed.

\section{Conclusions}
\label{sec:unified}

We have explored the broad emission line region (BELR) in over 30,000 $z>1.54$ SDSS quasars, concentrating on the properties of the \civ\ emission line.  We consider two well-known effects involving the \civ\ emission line: the anti-correlation between the \civ\ EQW and luminosity (i.e, the Baldwin Effect) and the blueshifting of the peak of \civ\ emission with respect to the systemic redshift.   Both of these effects are further considered in the context of the similarly well-known anti-correlation of the UV and X-ray continua (i.e., the $L_{\rm UV}$--$\alpha_{\rm ox}$ relationship).  Based on our investigations, we draw the following conclusions:
\begin{itemize}
\item Both radio-quiet (RQ) and radio-loud (RL) quasars show a Baldwin Effect anti-correlation between \civ\ EQW and UV luminosity (Figure~\ref{fig:beff1550}).
\item Similarly both RQ and RL have \civ\ emission lines that are shifted blueward of systemic.  Unlike the BEff, the blueshifts are very different for RQ and RL quasars (Figure~\ref{fig:c4bhist}).  For RQ quasars the median is $810\;{\rm km\,s^{-1}}$, while for RL quasars it is only $360\;{\rm km\,s^{-1}}$.  The blueshifts are luminosity dependent, with RL blueshifts having a weaker $L$ dependence (Figure~\ref{fig:c4b1550}).  Analysis of composite spectra (Figures~\ref{fig:lineplot3way} and \ref{fig:lineplot3waypanels}) suggest that the SED rather than $L$ is driving the observed differences --- consistent with the known $L_{\rm UV}$--$\alpha_{\rm ox}$ relationship.
\item In a 2-D \civ\ EQW vs.\ \civ\ blueshift parameter space (Figure~\ref{fig:c4beqw}), the combination of the above two results means that the quasar distribution has interesting structure.  The average UV emission-line properties of quasars in different regions of this parameter space indicate that there are substantive and systematic differences in the properties of the BELR across this space.
As such, we suggest that this space can be used to categorize quasars --- illuminating important physical differences among the population.
For example, strong BALQSOs and RL quasars largely live in opposite corners of \civ\ EQW-blueshift parameter space  (Figures~\ref{fig:c4beqwmorph} and \ref{fig:c3bhist}).
\item We conclude that these two \civ\ parameters are capturing an important trade-off between ``disk'' and ``wind'' components to UV emission lines that is ultimately sensitive to the shape of ionizing continuum.  More work is needed to probe the shape of the full ionizing continuum, but the ratio of UV to X-ray flux as measured by $\alpha_{\rm ox}$ captures the trends seen in the \civ\ EQW-blueshift parameter space with strong \civ\ EQW indicating a more ionizing SED and large \civ\ blueshift indicating a less ionizing SED (Figure~\ref{fig:wu09b}).
\item While RL quasars are predominantly found in the upper left corner of this parameter space, RQ quasars are also found there (Figure~\ref{fig:c4beqw}).  The relative fractions of these two populations suggests that there is a parent population of RQ quasars from which RL quasars are drawn.   Something other than the shape of the ionizing continuum (e.g., BH spin) must therefore be implicated in triggering radio jets.  Curiously, but in agreement with some previous results, orientation seems to have little influence on the location of RL quasars in the \civ\ parameter space (Figures~\ref{fig:c4beqwmorph}).
\item The differences seen across the \civ\ EQW-blueshift parameter space may have important implications for determining bolometric luminosities, BH masses, and Eddington ratios.  Specifically, it is important to determine if bolometric corrections are systematic across the \civ\ parameter space and if the core reverberation mapping sample results can be applied to objects that lie outside of the small parameter space that those objects probe (Figure~\ref{fig:kaspi}).
\end{itemize}

\section{Acknowledgments}

GTR acknowledges support from an Alfred P. Sloan Research Fellowship and NASA grant 07-ADP07-0035.  GTR thanks Daniel Proga and Michael Eracleous for fruitful discussions.  PCH acknowledges support from the STFC-funded Galaxy Formation and Evolution programme at the Institute of Astronomy.   PBH is supported by NSERC.  Funding for the SDSS and SDSS-II has been provided by the Alfred P. Sloan Foundation, the Participating Institutions, the National Science Foundation, the U.S. Department of Energy, the National Aeronautics and Space Administration, the Japanese Monbukagakusho, the Max Planck Society, and the Higher Education Funding Council for England. The SDSS is managed by the Astrophysical Research Consortium for the Participating Institutions. The Participating Institutions are the American Museum of Natural History, Astrophysical Institute Potsdam, University of Basel, Cambridge University, Case Western Reserve University, University of Chicago, Drexel University, Fermilab, the Institute for Advanced Study, the Japan Participation Group, Johns Hopkins University, the Joint Institute for Nuclear Astrophysics, the Kavli Institute for Particle Astrophysics and Cosmology, the Korean Scientist Group, the Chinese Academy of Sciences (LAMOST), Los Alamos National Laboratory, the Max-Planck-Institute for Astronomy (MPIA), the Max-Planck-Institute for Astrophysics (MPA), New Mexico State University, Ohio State University, University of Pittsburgh, University of Portsmouth, Princeton University, the United States Naval Observatory, and the University of Washington.

{\it Facilities:} \facility{Sloan}.

\section*{Appendix}
\label{sec:appendix}

A simplified version of the query that was used to merge the DR7 spectroscopic pipeline with the formal Schneider et al.\ (2010) DR7 quasar catalog is given as an example below. The actual input (public.gtr.DR7qsosHW10pub2) and output tables (public.gtr.dr7qsos\_c4b\_HW10pub2n\_big) are publicly available through the SDSS CasJobs server at http://casjobs.sdss.org/CasJobs/ .  The input table is the DR7 quasar catalog with the HW10 redshifts prepended; these can also be obtained from http://www.sdss.org/dr7/products/value\_added/ .

\begin{verbatim}
SELECT
q.ra,q.dec,
c.bestobjid,
c.specobjid,
c.ibest-q.extinction_i as i,
c.ibesterr,
q.extinction_i,
q.z as zSDSS,
c.zemHW,
c.zemHWalt,
c.delgi,
so.sn_2,
c.firstpeak,
isnull(f.integr,-1) as integr,
isnull(r.cps,-9) as xray,
sl1.lineid as CIVlineID,
sl1.restWave as CIVrestwave,
sl1.wave as CIVwave,
sl1.waveErr as CIVwaveErr,
sl1.sigma as CIVsigma,
sl1.sigmaErr as CIVsigmaErr,
sl1.height as CIVheight,
sl1.ew as CIVew,
sl1.ewerr as CIVewErr,
sl1.continuum as CIVcont,
sl2.lineid as MgIIlineid,
sl2.restWave as MgIIrestwave,
sl2.wave as MgIIwave,
sl2.waveErr as MgIIwaveErr,
sl2.sigma as MgIIsigma,
sl2.sigmaErr as MgIIsigmaErr,
sl2.height as MgIIheight,
sl2.ew as MgIIew,
sl2.ewerr as MgIIewErr,
sl2.continuum as MgIIcont,
(sl1.wave/1549.06)-1 as c4z,
(sl2.wave/2798.75)-1 as mg2z,
((c.zemHW)-((sl1.wave/1549.06)-1))*2.9979e5/(1.0+c.zemHW) as c4b,
((c.zemHW)-((sl3.wave/1908.73)-1))*2.9979e5/(1.0+c.zemHW) as c3b
INTO myDB.dr7qsos_c4b_HW10pub2n_big
FROM public.gtr.DR7qsosHW10pub2  as c
left outer join SpecPhotoAll as q on q.specObjID = c.specObjID
left outer join SpecLine sl1 on q.specObjID = sl1.specObjID
left outer join SpecLine sl2 on q.specObjID = sl2.specObjID
left outer join SpecLine sl3 on q.specObjID = sl3.specObjID
left outer join First as f on q.objID = f.objID
left outer join Rosat as r on q.objID = r.objID
left outer join SpecObjAll as so on q.specObjID = so.specObjID
WHERE (
q.z>=1.54 AND q.z<=4.5 AND
sl1.lineID=1549 AND sl2.lineID=2799
)
ORDER by q.ra
\end{verbatim}

\bibliography{ms}
\bibliographystyle{apj3}

\end{document}